\documentclass{aa}  
\newcommand{\degree}{\ensuremath{{}^{\circ}}\xspace}
\newcommand{\xb}{XB 1916-053 }

\bibpunct{(}{)}{;}{a}{}{,} 
\usepackage{natbib}
\usepackage{graphicx}
\usepackage{lscape}
\usepackage{rotating}
\usepackage{subfig}
\usepackage{placeins}
\usepackage{hyperref}
\usepackage{txfonts}

\begin{document} 

   \title{Spectral analysis of the dipping LMXB system XB 1916-053}
%
%
\author{A. F. Gambino\inst{\ref{inst1}} \and R. Iaria\inst{\ref{inst1}} 
\and T. Di Salvo\inst{\ref{inst1}} \and S. M. Mazzola\inst{\ref{inst1}} \and A. Marino\inst{\ref{inst1},\ref{inst4},\ref{inst5}} \and L. Burderi\inst{\ref{inst2}}\and A. Riggio\inst{\ref{inst2}} \and A. Sanna\inst{\ref{inst2}} \and \\ N. D'Amico\inst{\ref{inst3}}
 }

\institute{Universit\`a degli Studi di
  Palermo, Dipartimento di Fisica e Chimica, via Archirafi 36 - 90123 Palermo, Italy\label{inst1}\\
  \email{angelofrancesco.gambino@unipa.it}
                   \and
                   Universit\`a degli Studi di Cagliari, Dipartimento di Fisica, SP
           Monserrato-Sestu, KM 0.7, 09042 Monserrato, Italy\label{inst2}
           \and
           INAF-Osservatorio Astronomico di Cagliari, via della Scienza 5, 
           09047 Selargius, Italy\label{inst3}
           \and
           INAF/IASF Palermo, via Ugo La Malfa 153, I-90146 - Palermo, Italy\label{inst4}
          \and
           IRAP, Universit\`e de Toulouse, CNRS, UPS, CNES, Toulouse, France\label{inst5}
           }


 %
  \abstract
   {XB 1916-053 is a low mass X-ray binary system (LMXB) hosting a neutron star (NS) and showing periodic dips. The spectrum of the persistent emission was modeled with a blackbody component having a temperature between 1.31 and 1.67 keV and with a Comptonization component with an electron temperature of 9.4 keV and a photon index $\Gamma$ between 2.5 and 2.9. 
    The presence of absorption features associated with highly ionized elements suggested the presence of partially ionized plasma in the system. }
   {In this work we performed a study of the spectrum of XB 1916-053, which aims to shed light on the nature of the seed photons that contribute to the Comptonization component. 
   }
   {We analyzed three Suzaku observations of XB 1916-053: the first was performed in November 2006 and the others were carried out in October 2014. We extracted the persistent spectra from each observation and combined the spectra of the most recent observations, obtaining a single spectrum with a higher statistic. We also extracted and combined the spectra of the dips observed during the same observations.}
   {On the basis of the available data statistics, we infer that the scenario in which the corona Comptonizes photons emitted both by the innermost region of the accretion disk and the NS surface is not statistically relevant with respect to the case in which only photons emitted by the NS surface are Comptonized. We find that the source is in a soft spectral state in all the analyzed observations. We detect the K$\alpha$ absorption lines of \ion{Fe}{xxv} and \ion{Fe}{xxvi}, which have already been reported in literature, and for the first time the K$\beta$ absorption lines of the same ions. We also detect an edge at 0.876 keV, which is consistent with a \ion{O}{viii} K absorption edge. The dip spectrum is well described by a model that considers material in different ionization states covering the persistent spectrum and absorbing part of the rear radiation. From this model we rescale the distance of the absorber to a distance that is lower than 1$\times10^{10}$ cm.}
   {   }


\maketitle

\section{Introduction}
\label{sec:Intro}

XB 1916-053 is a low mass X-ray binary system (LMXB) showing dips and type I X-ray bursts. 
Type I X-ray bursts were observed for the first time by \cite{Becker_77} using \textit{OSO 8} data and revealed the presence of a neutron star (NS) in the binary system. The distance to the source was evaluated by several authors taking into account the X-ray bursts showing photospheric radius expansion (PRE), which are characterized by luminosities reaching the Eddington limit. \cite{Smale_88} evaluated a distance of 8.4 kpc for an accreted material with cosmic abundances of hydrogen, or else of 10.8 kpc for a hydrogen-deficient material.\\ \cite{Galloway} suggested a distance of 7 kpc or 9 kpc, for a H-rich and He-rich plasma accreted onto the NS surface, respectively. Moreover, \cite{Yoshida_PhD_93} suggested a distance of 9.3 kpc with a 15\% error, on the basis of the study of a type I X-ray burst that showed PRE observed with the \textit{Ginga} satellite.\\
\cite{Church_97} evaluated the orbital period of the system, taking advantage of the periodicity of the dips present in the \textit{ASCA} data. The dips are evidence of the partial covering of the central emission region of the binary system by a bulge of cold and/or partially ionized material in the point at which the accretion flow hits the outer part of the accretion disk. These authors estimated an orbital period of $P=3005 \pm 10 $ s. 
The observation in the optical band showed a modulation of the light curve with a period of 3027.4$\pm$ 0.4 s \citep{Grindlay_88}. The successive extension of the orbital ephemeris \citep{Iaria_15b} revealed that this modulation may be caused by the presence of a third body in the system. The mass of this body and its orbital period are dependent on the mass transfer scenario adopted for XB 1916-053. The authors evaluated a mass for the third body of $M_3=0.10 - 0.14$ M$_{\odot}$ and an orbital period of 51 yr assuming a conservative mass transfer, or $M_3=0.055$ M$_{\odot}$ and an orbital period of 26 yr if the mass transfer is nonconservative. In the latter case, the analysis also returned an estimation of the companion star mass ($M_2=0.028 $ M$_{\odot}$ ), that is a helium-rich star, according to \cite{nelemans}. The presence of the dips in the light curve suggests that the system is seen with a high inclination angle, i.e., between 60\degree and 80\degree \citep{Frank_1987}.\\
The spectrum of XB 1916-053 has been widely investigated both for the continuum and the contribution of the dips. \cite{zhang_2014}, using Suzaku data of November 2006, observed the source in a high-soft spectral state and modeled the continuum spectrum of XB 1916-053 taking into account a single corona that Comptonizes the photons emitted both by the NS surface and by the innermost part of the accretion disk. This model allowed these authors to constrain the spectral parameters of the accretion disk, as well as those related to the blackbody component associated with the emission of the NS surface. They obtained an equivalent hydrogen column density of $N_H=(6\pm1) \times 10^{20}$ cm$^{-2}$, inferring spectral parameters that are typical of systems in a soft spectral state. They found an inner disk temperature of $kT_{in} = 1.87\pm 0.18$ keV and a temperature and photon index for the electron corona of $kT_e=3.7^{+1.8}_{-0.4}$ keV and $\Gamma=2.65^{+1.28}_{-0.46}$, respectively. \\
The spectrum also shows some absorption lines superimposed on the continuum emission. The doublet of the \ion{Fe}{xxv} and \ion{Fe}{xxvi} K$\alpha$ absorption lines was detected for the first time by \cite{Boirin_xmm} with \textit{XMM-Newton} data. The authors also found weak evidence of the presence of the \ion{Mg}{xii}, \ion{S}{xvi}, \ion{and Ni}{xxvii} K$\alpha$ absorption lines, as well as the possible presence of the \ion{Fe}{xxvi} K$\beta$ absorption line and an absorption edge at 0.98 keV. The authors proposed that the latter should be a K absorption edge from moderately ionized Ne ions, or an L absorption edge from moderately ionized Fe ions. They also proposed the possibility that this should be the result of a superimposition of the edges of both the kind of ions.\\
\cite{Iaria_06}, using Chandra data, confirmed the presence of the \ion{Mg}{xii} and \ion{S}{xvi} K$\alpha$ absorption lines, and furthermore detected the K$\alpha$ absorption lines of \ion{Ne}{x} and \ion{Si}{xiv}. The diagnostic of the absorption lines with Chandra allowed the authors to obtain an estimation of the line widths of these narrow features and the following estimation of the emission region at which the lines originate: $ 4\times 10^{10}$ cm far away from the NS, that is almost at the accretion disk outer rim.\\
In \cite{Boirin} the transition from the persistent to the dip spectrum was explained as a decrease of the ionization parameter $\xi$ of the plasma with a simultaneous increase of the equivalent hydrogen column density of a ionized absorber interposed between the observer and the central emitting source. Furthermore, because of the great variety of absorption features into the spectrum of XB 1916-053, this absorber is likely to be partially ionized plasma.\\
The aim of this work is the detailed analysis of the continuum and absorption lines present in the spectrum of XB 1916-053, taking advantage, for the first time,  of all the available observations performed by \textit{Suzaku}.

The paper is structured as follows: In \autoref{sec:observation} we report the data selection and reduction, in \autoref{sec:analysis} we report the spectral analysis performed on the persistent and dip spectra, whilst we discuss the obtained results in \autoref{sec:discussion}. A conclusion is reported in \autoref{sec:conclusion}.

\section{Observation and data reduction}
\label{sec:observation}

\begin{figure*}
\centering

\includegraphics[angle=-90, width=8.5cm]{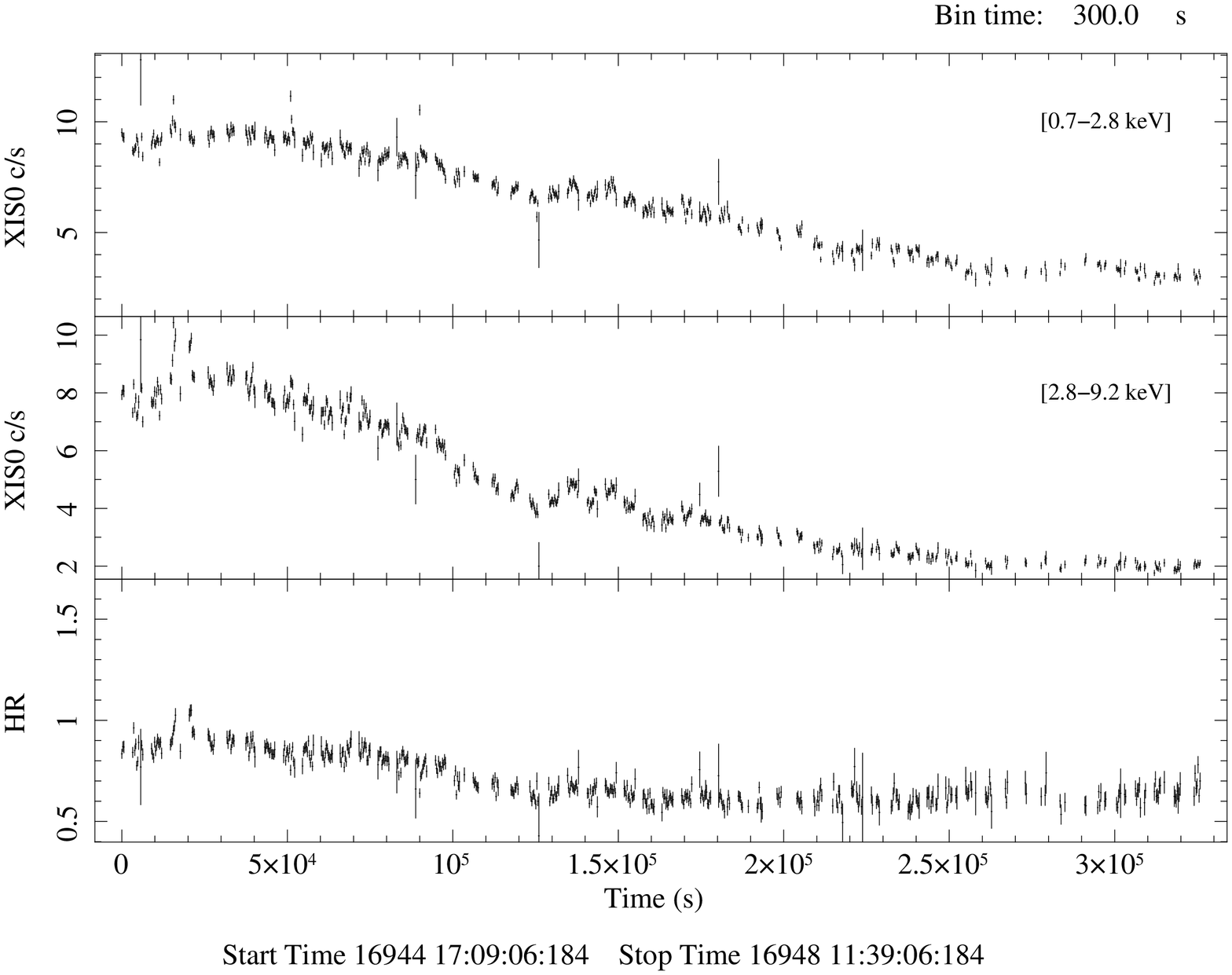}\hspace{0.2truecm}
\includegraphics[angle=-90, width=8.5cm]{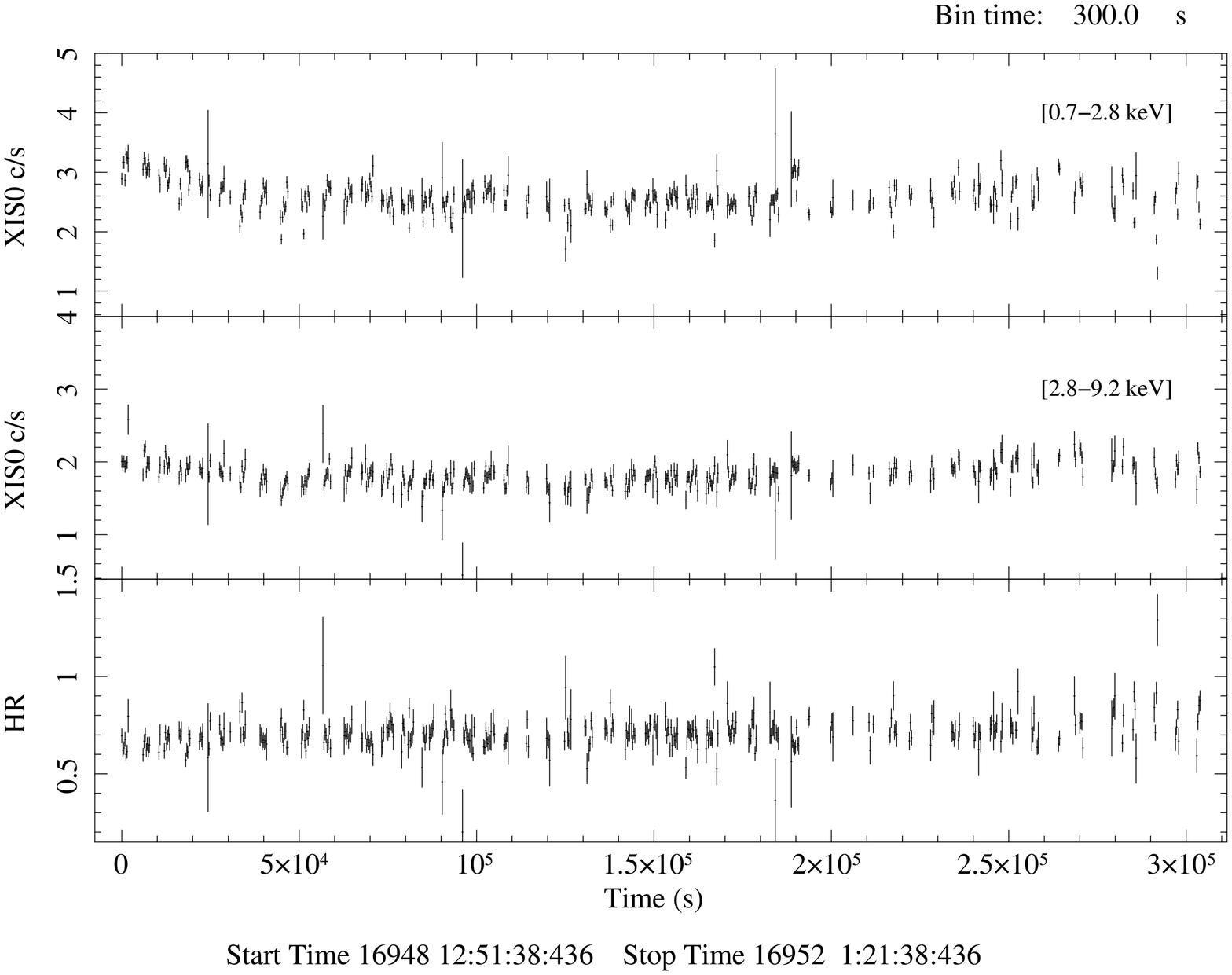}
\caption{Light curves from XIS0 in the energy bands 0.7-2.8 keV (top panels) and 2.8-9.2 keV (middle panels) for ObsID 409032010 (Obs 2, left) and  ObsID 409032020 (Obs 3, right), respectively. The corresponding hardness ratios are reported in the bottom panels. }
\label{fig:HR_persistent_XIS0}
 \end{figure*}

Suzaku \citep[][]{Mitsuda} observed XB 1916-053 three times. The first observation (ObsID 401095010, hereafter Obs 1) was performed from 8 November 2006 at 05:49:53 to 9 November 2006 at 02:41:15, for a net exposure of about 39 ks. The other two observations were acquired in 2014 and in particular the first observation (ObsID 409032010, hereafter Obs 2) was performed from 14 October   at 16:20:52 to 18 October at 12:20:16, for a net exposure of 155.9 ks. The second observation (ObsID 409032020, hereafter Obs3) was performed from 18 October at 12:20:17 to 22 October at 02:40:09, for a net exposure of 140.5 ks.

During all the observations both the X-ray Imaging Spectrometers \citep[XIS;][]{Koyama} and the Hard X-ray Detector \citep[HXD;][]{Taka} were operative. 
The XIS detectors consist of four chips generally numbered from 0 to 3 that are sensible to the energy band between 0.2 and 12 keV. The XIS0, XIS2, and XIS3 have a similar response to the incoming radiation, consisting of the same kind of front-illuminated charge-coupled device (CCD). On the other hand, the XIS1 uses a back-illuminated CCD.
Unfortunately, the XIS2 broke down in 2006, moreover during Obs 1, and it has shut down ever since. Therefore the only existing XIS2 data in our sample are relative to a fraction of Obs 1.
The HXD assembly, on the other hand, consists of two nonimaging detectors: the p-i-n type Si photo diode (PIN) that is sensible to the energy band 10--70 keV and the Gd$_2$SiO$_5$(Ce) scintillators (GSO) that are sensible to the energy range 30--600 keV.\\
In Obs 1 the XIS chips operated in full window mode and collected data edited both in 3x3 and 5x5 pixel mode for a net exposure of 40 ks, while the HXD-PIN detector was exposed for 37 ks. \\
On the other hand, in Obs 2, the XIS0, XIS1, and XIS3 detectors operated using the 1/4 window option and collected data edited both in 3x3 and 5x5 pixel mode for a net exposure of 155.9 ks, while the HXD-PIN detector was exposed for 8.8 ks. 
Similar to the previous observation, in Obs 3 all the XIS detectors operated using the 1/4 window option and collected data edited both in 3x3 and 5x5 pixel mode for a net exposure of 140.5 ks, whilst the HXD-PIN collected data for 6.9 ks.\\
The source was not detected in the GSO, and for this reason we limited our analysis to the spectra collected by the XIS and PIN detectors for all the observations.
The data were analyzed inside the HEASoft v. 6.19 environment. Moreover, for Obs 1 we used the calibration files v. 20070731 and v. 20070710 for XIS and HXD/PIN, respectively. Similarly, for Obs 2 and Obs 3 we used the XIS calibration files v. 20160607 and the HXD-PIN calibration files v. 20110913.\\
All the data were processed with the \verb|aepipeline| routine provided with the Heasoft FTOOLS.  For each XIS chip we combined the data edited in 3x3 and 5x5 pixel format and performed a better estimation of the attitude using the \verb|aeattcor.sl| tool created and distributed by J. E. Davis. The attitude correction to the XIS data was applied with the supplementary tool \verb|xiscoord|.

The light curve of Obs 1 showed an increasing trend in the count rate from about 5 c/s up to 11 c/s. The observation showed periodic dips and one type I X-ray burst, as already observed by \cite{zhang_2014}.\\
Instead, the Obs 2 initial count rate of about 18 c/s decreased until a final value of about 5 c/s. Four type I X-ray bursts are present in the observation as well as several periodic dips. The hardness ratio of the source during the observation, however, did not show significant variations and maintained an approximately constant trend.
On the other hand, during Obs 3 the source maintained an approximately constant count rate of about 5 c/s and several periodic dips are present. Also in this case, the hardness ratio did not show any important variations.\\
In this work we were not interested in the analysis of the type I X-ray bursts; we then removed a 5 s time segment before the rise time of each burst and a time interval 100 s long subsequent the onset of the burst.
Furthermore, we distinguished between the dip and the persistent emission, performing a time selection on the data of all the observations.\\
To estimate if the XIS data were affected by pileup, we used the tool \verb|pile-estimate.sl| created by M. A. Novak. 
For Obs 1 we obtained a pileup fraction between 9\% and 13\% for all the XIS chips considering the events in a circular region centered at the coordinates of XB 1916-053 \citep[R.A.= 289.699462\degree and Dec.= -5.238081\degree (J2000);][]{Iaria_06} and with radius 114".
Then, in order to reduce the pileup fraction under the 4\% threshold, i.e., the tolerance limit we assumed in this work, we only considered the events included in an annular region centered onto the source and having an outer radius of 114" and an inner radius of 31". On the other hand, for Obs 2 and Obs 3 a circular region centered at the coordinates of the source and having a radius of 114" is sufficient to obtain an estimation of the pileup fraction of the 4\% at most for Obs 2 and of about 2\% for Obs 3. These two observations, indeed, were acquired using the 1/4 windows option that allows us to increase the window frame time and to observe the source without pileup also to higher count rates with respect to the full window mode. 
The XIS spectra of the source for each observation were obtained considering events falling inside these extraction regions. The corresponding background spectra were extracted selecting events inside a circular region localized far away from the source and covering the same area of the corresponding source extraction region. 
For Obs 1 we extracted the spectra of the persistent emission and of the dips taking into account the data collected by each XIS chip. For the XIS2 data, we excluded the last 6 ks of observation during which the instrument did not acquired data anymore, owing to the damage it had undergone. As a consequence of this, we took into account only the spectra extracted from the XIS0, XIS1, and XIS3 data for Obs 2 and Obs 3.\\ 
The response of each XIS chip was obtained using the tool \verb|xisrmfgen|, whilst the ancillary response files (ARFs) were obtained using the tool \verb|xissimarfgen|, setting as coordinates for the source R.A.= 289.699462\degree and Dec.= -5.238081\degree \citep[][]{Iaria_06}.\\
The HXD-PIN spectra were extracted using the tool \verb|hxdpinxbpi|, while we used the response files provided by the HXD team. We selected the response files {\tt ae\_hxd\_pinxinome3\_20080129.rsp} for Obs 1 and {\tt ae\_hxd\_pinxinome11\_20110601.rsp} for Obs 2 and Obs 3, respectively.\\
The background file of Obs 1 was regularly produced by the HXD team and then used in the analysis, whilst the background files for the Obs 2 and Obs 3 of 2014 were not available.
Starting from that year, indeed, normal operation of the HXD was very rare owing to the power consumption of the spacecraft. In order to estimate the background generated by the NXB+CXB (Non-X-ray Background + Cosmic X-ray Background) events, the HXD team suggested that we use the spectrum of the source RXJ 1856.5-3754, that is a NS with a considerably soft X-ray emission, where the PIN data are almost attributable to the CXB + NXB contribution. Since our source is very bright with respect to RXJ 1856.5-3754, the background systematic error is not so significant below 30 keV, whilst the NXB systematics above 30 keV may be around the 10-20\% in Obs 2 and Obs 3.\\
Then, we processed the ObsID 109008010 data of RXJ 1856.5-3754 of 2014 April with the \verb|aepipeline| routine. This observation has an exposure of 35 ks that has been considered in its entirety to extract the HXD/PIN spectrum of the source through the \verb|hxdpinxbpi| tool. 
To take into account a systematic error of the 20\% above 30 keV, we used the ftool \verb|grppha|. Hereafter, this spectrum represents the  HXD/PIN CXB + NXB background spectrum for both Obs 2 and Obs 3. \\
The products of this first processing step for Obs 1 are spectra of the persistent emission of XB 1916-053 for a net exposure of 33 ks for the XIS0 and XIS3, and  30 ks for the XIS2. Furthermore, the corresponding HXD/PIN spectrum has an exposure of 30 ks.\\
From Obs 2 we extracted XIS persistent spectra of 120 ks of exposure for each considered chip respectively and a HXD/PIN persistent spectrum of 8 ks of exposure. Furthermore, from this observation we also extracted XIS spectra of the dips for an amount of 13 ks of exposure for each XIS chip, as well as a HXD/PIN spectrum of the dips of 502 s exposure.
On the other hand, for the Obs 3 we extracted XIS spectra of the persistent emission for an amount of 83 ks of exposure for each of the considered chips, and of 1.5 ks of exposure from the HXD/PIN data. With regard to the dip part, we extrapolated spectra of 23 ks from the XIS data of each considered chip, as well as a HXD/PIN spectrum of 5 ks of exposure.\\
Considering the similar spectral response of XIS0, XIS2, and XIS3, and once we tested the actual compatibility of the spectra acquired by these instruments, in order to increase the statistics we chose to combine the spectra extracted from these devices and their spectral responses using the \verb|addascaspec| routine. We did this for all the observations and both for the persistent and the dip spectra; we remark that during the Obs 2 and Obs 3, only the XIS0 and XIS3 were active.\\
Furthermore, we noticed that the spectral shape of the persistent emission of XB 1916-053 was the same in Obs 2 and Obs 3. Then, since these observations are extremely close in time, we combined the XIS0 and XIS3 spectra (hereafter XIS03 spectrum) and each HXD/PIN spectrum of the two observations to further increase the statistic of the observations. We obtained a broadband spectrum with a considerably high statistics: the resulting exposure times for the XIS03 and PIN total combined spectra are of 400 ks and 9.3 ks, respectively.

\section{Data analysis}
\label{sec:analysis}

\subsection{Persistent spectrum}

\begin{table*}
\caption{Results of the fits on Spectrum 1 and Spectrum 2 adopting the 1-seed and 2-seed models, respectively. The errors for each spectral parameter are quoted at the 90 \% of statistical confidence. }
\footnotesize
\begin{tabular}{ll|cc|cc}
\hline
 \multicolumn{2}{c}{    } & \multicolumn{2}{c}{ Model: 1-seed} & \multicolumn{2}{c}{Model: 2-seed}  \\
 \hline

 Component & Parameter & Spectrum 1 & Spectrum 2 & Spectrum 1 & Spectrum 2 \\
 \hline 
{\sc phabs} & N$_H$ (10$^{22}$)&  0.433$^{+0.012}_{-0.010}$ & 0.479$^{+0.014}_{-0.013}$ & 0.45$^{+0.03}_{-0.02}$ & 0.479$^{+0.020}_{-0.010}$ \\
 
{\sc gabs}  & LineE (keV)  & -- & 4.4$\pm$0.3  & -- & 4.4$^{+0.2}_{-0.3}$ \\
                                 & Sigma(keV)  & -- & 1.4$^{+0.4}_{-0.2}$ & -- & 1.4$^{+0.3}_{-0.3}$ \\
                                 & Strength       & -- & 0.7$^{+1}_{-0.4}$ & -- & 0.7$^{+0.7}_{-0.2}$  \\
                                 
{\sc edge}  & edgeE (keV) & -- & 0.876$^{+0.02}_{-0.013}$ & -- & 0.876$^{+0.014}_{-0.013}$ \\
                                 & MaxTau           & -- & 0.11$\pm$0.02 & -- &  0.11$\pm$0.02  \\
 
{\sc bbody} & kT(keV) & $1.3^{+0.2}_{-0.1}$ & $1.1^{+0.1}_{-0.3}$ & $1.9^{+0.2}_{-0.4}$ & $1.11^{+0.03}_{-0.2}$ \\
                               &  Norm. (10$^{-3}$)  &  < 0.95 & 1$^{+2}_{-1}$ & < 1.44 & <3.5 \\
                                           
                                   &    {\sc bbody Unabs. Flux} (erg cm$^2$ s$^{-1}$)  & <8$\times 10^{-11}$ &  (1.2$\pm$0.1)$\times 10^{-10}$  & <1$\times 10^{-10}$   &  <3$\times 10^{-10}$  \\                                     
                                           
{\sc diskbb}     &  T$_{in}$ (keV) &     $0.69\pm 0.03$   & 0.55$^{+0.05}_{-0.04}$  &      0.60$^{+0.07}_{-0.1}$ & 0.55$^{+0.05}_{-0.04}$ \\
                                           &  $(R_{in}/D_{10})^{2}\; cos(\theta)$ (km$^2$ kpc$^{-2}$) & 44$\pm 7$ & 68$^{+25}_{-19}$ & <33 & 68$^{+20}_{-19}$ \\
                                           &    {\sc diskbb Unabs. Flux} (erg cm$^2$ s$^{-1}$)  & (2.0$\pm$0.2)$\times 10^{-10}$ &  (1.3$\pm$0.1)$\times 10^{-10}$  & <$9\times 10^{-11}$   &  (1.3$\pm$0.1)$\times 10^{-10}$  \\                                            

{\sc NthComp} &  $\Gamma$ & $2.6^{+0.3}_{-0.5}$ & $1.5^{+0.5}_{-0.3}$ &  $2.2^{+0.7}_{-0.3}$  &  $1.5^{+0.5}_{-0.3}$ \\
                                           &   kT$_{e}$ (keV) & >7 &  $5^{+3}_{-1}$ &   $\geq 1.2 $ &  $5^{+2}_{-1}$ \\
                                           &   kT$_{bb}$ (keV) & $1.3^{+0.2}_{-0.1}$ &  $1.10^{+0.09}_{-0.3}$ &  $1.9^{+0.2}_{-0.4}$  &  $1.11^{+0.03}_{-0.2}$ \\
                                           &   inp\_type (0/1) & 0 & 0& 0 & 0 \\
                                           &   Redshift &  0 & 0 &  0 &  0 \\
                                           &   Norm. (10$^{-3}$) &  $5^{+1}_{-3}$ & $2^{+5}_{-1}$ &  < 3.4 & < 11.8 \\
                                           
                                           &    {\sc  NthComp Unabs. Flux} (erg cm$^2$ s$^{-1}$)  & (3.8$\pm$0.4)$\times 10^{-10}$ &  (2.9$\pm$0.3)$\times 10^{-10}$  & <3$\times 10^{-10}$   &  <4$\times 10^{-10}$  \\ 
                                           
{\sc NthComp} &  $\Gamma$ & -- & -- &  $2.2^{+0.7}_{-0.3}$  &  $1.5^{+1}_{-0.3}$ \\
                                           &   kT$_{e}$ (keV) & -- &  -- &  >1 &  $5^{+2}_{-1}$ \\
                                           &   kT$_{bb}$ (keV) & -- &  -- & $0.60^{+0.07}_{-0.1}$ &  $0.55^{+0.05}_{-0.04}$ \\
                                           &   inp\_type (0/1) & -- & -- & 1 & 1 \\
                                           &   Redshift &  -- & -- &  0 &  0 \\
                                           &   Norm. (10$^{-3}$) &  -- & -- & $41^{+22}_{-21}$  & <31  \\ 
                                           
                                           &    {\sc  NthComp Unabs. Flux} (erg cm$^2$ s$^{-1}$)  & -- &  --  & (3.4$\pm$0.3)$\times 10^{-10}$   &  <4$\times 10^{-10}$  \\                         
                                           
{\sc Gaussian} &   LineE (keV) & 6.65$\pm$0.03 & 6.69$^{+0.03}_{-0.02}$ & 6.65$\pm$0.03 & 6.69$\pm$0.02 \\
(\ion{Fe}{xxv} K$\alpha$)                                  &   Sigma (keV) &  0.02 (frozen) & 0.02 (frozen) & 0.02 (frozen) & 0.02 (frozen) \\
                                           &   Norm (10$^{-5}$) & -3$\pm$1 & -3.9$^{+0.6}_{-0.7}$ & -3$\pm$1 & -3.9$\pm$0.7  \\
                                           &   {\sc Detection significance} ($\sigma$)&  4  & 9 & 4 & 10 \\
                                           &  {\sc Eq. Width } ($keV$) & -0.011$\pm$0.003 & -0.005$\pm$0.002  &  -0.014$\pm$0.003 &  -0.013$\pm$0.002   \\
                                            
{\sc Gaussian} &   LineE (keV)  & $6.943^{+0.02}_{-0.013}$ & $6.972^{+0.014}_{-0.010}$ & $6.943^{+0.02}_{-0.013}$ &  $6.974^{+0.012}_{-0.011}$  \\
(\ion{Fe}{xxvi} K$\alpha$)                                         &   Sigma (keV) &  0.02 (frozen) & 0.02 (frozen) & 0.02 (frozen) & 0.02 (frozen) \\
                                           &   Norm (10$^{-5}$) &  $-8\pm1$ & $-6.8\pm0.7$ & $-8\pm1$ & $-6.8^{+0.7}_{-0.6}$  \\    
                                           &   {\sc Detection significance} ($\sigma$)& 11  &  17 &  10 &  17  \\
                                           &  {\sc Eq. Width } ($keV$) & -0.029$\pm$0.003 & -0.017$\pm$0.002  &  -0.033$\pm$0.003 &  -0.026$\pm$0.002   \\               
                                           
{\sc Gaussian} &   LineE (keV)  & -- & 7.79$\pm$0.08 & -- & 7.79$^{+0.08}_{-0.04}$\\
(\ion{Fe}{xxv} K$\beta$)                                           &   Sigma (keV) &  0.02 (frozen) & 0.02 (frozen) & 0.02 (frozen) & 0.02 (frozen) \\
                                           &   Norm (10$^{-5}$) & -- & -1.3$\pm$0.7  & -- & -1.3$^{+0.6}_{-0.4}$ \\
                                           &   {\sc Detection significance} ($\sigma$)& -- &  3  & --  &  4  \\
                                           &  {\sc Eq. Width } ($keV$) & -- & -0.005$\pm$0.003  &  -- &  -0.006$\pm$0.003   \\                      

{\sc Gaussian} &   LineE (keV)  & -- & 8.19$\pm$0.07 & -- & 8.19$\pm$0.07\\
(\ion{Fe}{xxvi} K$\beta$)                                          &   Sigma (keV) &  0.02 (frozen) & 0.02 (frozen) & 0.02 (frozen) & 0.02 (frozen) \\
                                           &   Norm (10$^{-5}$) & -- & -1.7$\pm$0.7  & -- & -1.7$\pm$0.7  \\        
                                           &   {\sc Detection significance} ($\sigma$)& --  &  4  &  -- &  4  \\
                                           &  {\sc Eq. Width } ($keV$) & -- & -0.005$\pm$0.003  &  -- &  -0.010$\pm$0.003   \\                              

\hline 
                                           &    {\sc Total Unabs. Flux} ($\times 10^{-10}$ erg cm$^2$ s$^{-1}$)  & 5.8$\pm$0.6 &  5.3$\pm$0.5 & 5.8$\pm$0.6  &  5.4$\pm$0.5 \\
                                           &    $\chi^2/dof$ &  702.5/617  &   693.6/600  &  698.0/616  &  693.5/599  \\

\end{tabular}
\label{tab1}
\end{table*}

\begin{table*}
\caption{Results of the fit on Spectrum 1 and Spectrum 2, adopting the 1-seed and 2-seed models together with a component that takes into account the contribution of a local warm absorber, respectively. The errors for each spectral parameter are quoted at 90\% of statistical confidence. }
\footnotesize
\begin{tabular}{ll|cc|cc}
\hline
 \multicolumn{2}{c}{    } & \multicolumn{2}{c}{ Model: warm\_1seed} & \multicolumn{2}{c}{Model: warm\_2seed}  \\
 \hline
 Component & Parameter & Spectrum 1 & Spectrum 2 & Spectrum 1 & Spectrum 2 \\
 \hline 
{\sc phabs} & N$_H$ (10$^{22}$)&  0.432$^{+0.007}_{-0.009}$ & 0.47$\pm$0.02 & 0.46$\pm$0.02 & 0.47$\pm$0.02 \\

{\sc zxipcf}  & N$_H$ (10$^{22}$)  & 160$\pm$120 & 13$^{+8}_{-3}$  & 140$^{+130}_{-110}$ & 13$^{+8}_{-3}$ \\
                                 & Log($\xi$)  & 5.2$^{+0.4}_{-0.5}$ & 4.35$^{+0.20}_{-0.04}$ & 5.1$\pm$0.5 & 4.35$^{+0.2}_{-0.04}$ \\
                                 & CvrFract       & >0.87 & 1 & >0.87 & >0.97  \\
                                 & Redshift       & 0 & 0  & 0  & 0  \\
                                 
{\sc gabs}  & LineE (keV)  & -- & 4.38$^{+0.12}_{-0.11}$  & -- & 4.38$^{+0.11}_{-0.10}$ \\
                                 & Sigma(keV)  & -- & 0.72$^{+0.2}_{-0.14}$ & -- & 0.72$\pm$0.2 \\
                                 & Strength       & -- & 0.07$^{+0.05}_{-0.03}$ & -- & 0.07$^{+0.04}_{-0.02}$  \\
                                 
{\sc edge}  & edgeE (keV) & -- & 0.870$^{+0.014}_{-0.012}$ & -- & 0.869$^{+0.014}_{-0.012}$ \\
                                 & MaxTau           & -- & 0.11$^{+0.03}_{-0.02}$ & -- &  0.113$^{+0.03}_{-0.02}$  \\
 
{\sc bbody} & kT(keV) & $1.12^{+0.05}_{-0.11}$ & $0.750^{+0.20}_{-0.010}$ & 2.0$\pm$0.3 & $0.72^{+0.7}_{-0.06}$ \\
                               &  Norm. (10$^{-5}$)  &  < 45 & $\leq$ 50&  $\leq$128 & $\leq$20 \\
                                           
                                   &    {\sc bbody Unabs. Flux} (erg cm$^2$ s$^{-1}$)  & <4$\times 10^{-11}$ &  <4$\times 10^{-11}$  & <1$\times 10^{-10}$   &  <2$\times 10^{-11}$  \\                                       
                                           
{\sc diskbb}     &  T$_{in}$ (keV) &     $0.69^{+0.04}_{-0.05}$   & 0.51$^{+0.04}_{-0.06}$  &      0.52$^{+0.11}_{-0.6}$ & 0.50$\pm$0.05 \\
                                           &  $(R_{in}/D_{10})^{2}\; cos(\theta)$ (km$^2$ kpc$^{-2}$) & 47$^{+13}_{-7}$ & 78$^{+34}_{-13}$ & <28 & 82$^{+270}_{-9}$ \\
                                           &  {\sc diskbb Unabs. Flux} (erg cm$^2$ s$^{-1}$)  & (2.0$\pm$0.2)$\times 10^{-10}$ &  (1.1$\pm$0.1)$\times 10^{-10}$  & <6$\times 10^{-11}$   &  (1.1$\pm$0.1)$\times 10^{-10}$  \\                                            

{\sc NthComp} &  $\Gamma$ & $2.1^{+0.3}_{-0.2}$ & $1.84^{+0.04}_{-0.05}$ &  2.4$\pm$0.5  &  1.8$\pm$0.4 \\
                                           &   kT$_{e}$ (keV) & 7$\pm$1 &  6$\pm$1  &  12$^{+27}_{-9}$ &  6$\pm$1 \\
                                           &   kT$_{bb}$ (keV) & $1.12^{+0.05}_{-0.11}$ &  $0.750^{+0.2}_{-0.010}$ &  2.0$\pm$0.3  &  0.72$^{+0.7}_{-0.06}$ \\
                                           &   inp\_type (0/1) & 0 & 0& 0 & 0 \\
                                           &   Redshift &  0 & 0 &  0 &  0 \\
                                           &   Norm. (10$^{-3}$) &  $6.0^{+2}_{-0.5}$ & $10^{+12}_{-3}$ &  < 2.1 &  11.2$^{+3}_{-7}$ \\
                                           
                                           &    {\sc  NthComp Unabs. Flux} (erg cm$^2$ s$^{-1}$)  & (4.0$\pm$0.4)$\times 10^{-10}$ &   (4.0$\pm$0.4)$\times 10^{-10}$  & <3$\times 10^{-10}$   &  (4.0$\pm$0.4)$\times 10^{-10}$  \\ 
                                           
{\sc NthComp} &  $\Gamma$ & -- & -- &  2.4$\pm$0.5  &  1.8$\pm$0.4 \\
                                           &   kT$_{e}$ (keV) & -- &  -- &  12$^{+27}_{-9}$ &  6$\pm$1 \\
                                           &   kT$_{bb}$ (keV) & -- &  -- & 0.52$^{+0.11}_{-0.6}$ & 0.50$\pm$0.05 \\
                                           &   inp\_type (0/1) & -- & -- & 1 & 1 \\
                                           &   Redshift &  -- & -- &  0 &  0 \\
                                           &   Norm. (10$^{-3}$) &  -- & -- & $61^{+6}_{-21}$  & <8  \\ 
                                           
                                           &    {\sc  NthComp Unabs. Flux} (erg cm$^2$ s$^{-1}$)  & -- &  --  & (4.0$\pm$0.4)$\times 10^{-10}$   &  <9$\times 10^{-11}$  \\

\hline 
                                           &    {\sc Total Unabs. Flux} ($\times 10^{-10}$ erg cm$^2$ s$^{-1}$)  & 6.0$\pm$0.6 &  5.1$\pm$0.5 & 6.0$\pm$0.6  &  5.1$\pm$0.5 \\
                                           &    $\chi^2/dof$ &  698.30/618  &   707.51/605  &  691.30/617  &  707.48/604  \\

\end{tabular}
\label{tab2}
\end{table*}

The broadband spectrum obtained from Obs 1, hereafter named \textit{Spectrum 1}, covers the energy range between 0.8 keV and 40 keV. Similarly,  the spectrum obtained from the combination of data of Obs 2 and Obs 3 is named \textit{Spectrum 2} and covers the energy range between 0.8 keV and 30 keV.  
In these two spectra, however, the XIS data cover the energy range between 0.8 keV and 10 keV, whereas the data of the HXD/PIN cover the energy range between 15 keV and 40 keV in Spectrum 1 and between 15 KeV and 30 keV in Spectrum 2, respectively.
We ignored the energy ranges 1.7--1.9 keV and 2.2--2.4 keV in all the XIS spectra to exclude calibration uncertainties due to the presence of the instrumental K-edge of silicon and M-edge of gold, respectively.
In order to avoid an oversampling of the energy resolution of the XIS, we applied a grouping of a factor 4 to the data. Furthermore, we grouped all the spectra to have at least 25 photons per energy channel and to use the $\chi^{2}$ as an estimator of the goodness of the fit. The only exception to this is for the HXD/PIN data of Spectrum 1, where because of the lower statistics we grouped the spectrum to have at least 100 photons per energy channel.\\
In the subsequent part of this work we want to use the model adopted by \cite{zhang_2014} to describe the persistent spectrum of \xb, taking the advantage of the whole available Suzaku dataset. We seek to study the relative contribution of each spectral component with respect to the total observed flux, and we also compare this model with a simpler model that takes into account a single source of photons for the Comptonization that is represented by photons emitted by the NS surface. Moreover, we aim to study the emission features that are visible in the available spectra, trying to characterize the chemical abundances of the species from which they are produced.
To complete this task, as a first step we decided to adopt the simpler model to describe the continuum emission of the source.

For the spectral analysis we used Xspec v. 12.9.1 and the cross-section table of \cite{Verner} and the chemical abundances reported by \cite{Wilms}.  \\
We first fitted Spectrum 1 with a spectral model that is usually suitable for atoll class LMXB systems \citep[see, e.g.,][]{DiSalvo_2009, Piraino}. This model consists of a soft blackbody component plus a multicolor-disk blackbody and a hard component usually modeled with a Comptonization spectrum.\\
More precisely, to take into account the contribution of emission by the NS surface we used the {\tt bbody} model that is defined by a temperature $kT$ and a normalization. The disk contribution was modeled using the multicolor-disk blackbody component \citep[{\tt diskbb}, see][]{Mitsuda1984} that consists of two parameters: the temperature at the inner disk radius and a normalization parameter $N$ linked to the value of the inner radius through the relation $N=(R_{in}/D_{10})^2 \; cos(\theta)$, where $R_{in}$ is an apparent inner disk radius in km, $D_{10}$ is the distance to the source in units of 10 kpc, and $\theta$ is the inclination angle of the disk with respect to the line of sight ($\theta=0$ is the face-on case).
The contribution of a thermally Comptonized continuum was included by adopting the model {\tt nthComp} \citep[][]{Zycky}. This is defined by an asymptotic power-law photon index $\Gamma$ and by the temperatures $kT_e$ and $kT_{bb}$ of the electron cloud and seed photons, respectively.  
This model also takes into account a $redshift$ parameter and the parameter $inp\_type$ that can be set to 0 or 1, depending on the seed photons distribution as a blackbody or disk blackbody, respectively. As a first step, we assume that the photons that are subsequently Comptonized are those emitted by the NS surface, imposing $inp\_type$=0.
On the other hand, to take into account the photoelectric absorption by the neutral matter in the interstellar medium (ISM), we used the multiplicative spectral component \verb|phabs|, defined through the parameter $N_H$ that is the equivalent hydrogen column density (in units of $10^{22}$ atoms cm$^{-2}$). To take into account the different instrumental responses of the XIS and HXD/PIN detectors we also introduced a multiplicative intercalibration constant into the model.

The application of this model (hereafter 1-seed model) to Spectrum 1 returns a $\chi^2(d.o.f.)$ of 818.11(621) and evident residuals in absorption at about 7 keV, as observed by \cite{zhang_2014}. We modeled these residuals with the two absorption lines already found by \cite{Boirin_xmm} using the additive spectral component \verb|gauss| in Xspec. We fixed the value of the line widths to $\sigma$=20 eV due to the moderate energy resolution of the \textit{Suzaku}/XIS instrument. This value is compatible with the line width of the \ion{Fe}{XXVI} K$\alpha$ absorption line evaluated by \cite{Iaria_06} using data of the \textit{Chandra} observatory, and this value has been used for all the subsequent analysis.\\
The fit returned a value of $\chi^2(d.o.f.)$ of 702.5(617) and an F-test probability of chance improvement of $1.7\times10^{-19}$ with respect to the previous model.
We obtained two lines centered at 6.65 and 6.943 keV, actually compatible with the K$\alpha$ absorption lines of \ion{Fe}{xxv} and \ion{Fe}{xxvi}, respectively. The detection of these absorption lines was obtained with a confidence level of 4$\sigma$ and 11$\sigma$, respectively, and the obtained fit parameters are reported in Tab. \ref{tab1}.\\                                                      
Finally, from this best-fit model we extrapolated an unabsorbed flux of $(5.8\pm0.6)\times 10^{-10}$ erg cm$^{-2}$ s$^{-1}$ in the 0.1-100 keV energy band. The {\tt bbody} component provides a marginal contribution to the total flux since it contributes with less than the 14\% with respect to the total. On the other hand, the spectral components {\tt diskbb} and {\tt nthComp} contribute to the 35\% and 65\%, respectively.

In order to understand if substantial changes in the spectral shape of the source occurred between the observations of 2006 and 2014, we fitted Spectrum 2 with the same model. The fit returned a value of $\chi^2(d.o.f.)$ of 693(600) and evident residuals in absorption at about 8 keV, probably owing to the combination of more than one absorption line. The residuals also show a bump at about 4.5 keV and the clear presence of an edge at about 0.8 keV, that is the energy at which the K absorption edge of \ion{O}{viii} (E$\sim$ 0.871 keV) is expected. Then, we modeled this last feature with an absorption edge at 0.871 keV, while the absorption feature at about 8 keV was modeled with two Gaussian absorption lines with a fixed width of $\sigma$=20 eV. 
The bump in the residuals at 4.5 keV is often observed in high statistics Suzaku/XIS spectra and it is probably of systematic and nonphysical nature. To get rid of this feature, we modeled it with the spectral component {\tt gabs}. \\
The detection of the two K$\alpha$ lines remains confirmed as in Spectrum 1, but we find that they are stronger in Spectrum 2 because of its higher statistics. Indeed, the spectral model returned a detection of the \ion{Fe}{xxv} and \ion{Fe}{xxvi} K$\alpha$ lines with a level of confidence of about 9$\sigma$ and 17$\sigma$, respectively. 
In addition, the two supplementary absorption lines added returned energies that are consistent with those we expect for the K$\beta$ absorption lines of \ion{Fe}{xxv} and \ion{Fe}{xxvi}.  
Our fit allowed us to detect these lines with a level of confidence of about 3$\sigma$ and of 4$\sigma$, respectively.\\
At last, the model returned a value of unabsorbed flux of $(5.3\pm0.5)\times 10^{-10}$ erg cm$^{-2}$ s$^{-1}$ between 0.1 and 100 keV, to which contribute the {\tt bbody} component for about the 22\%, the {\tt diskbb} component  for the 24\%, and the {\tt nthcomp} component for about the 55\% of the total flux.\\
The results obtained from the application of this simple model both to Spectrum 1 and Spectrum 2 are typical of sources in a soft spectral state. Adopting a distance to the source of 9.3 kpc \citep[][]{Yoshida_PhD_93}, we obtained that the luminosity of XB 1916-053 is about the 2\% of the Eddington luminosity. According to this, taking into account the color correction factor $\psi$= 1.7 of \cite{Shimura_95} we obtained the inner radius of the accretion disk as $R_{disk}=\sqrt{norm/cos(\theta)}\;\psi^2 \; D_{10}$, where $norm$ is the normalization parameter of the {\tt diskbb} component, $\theta$ is the inclination angle of the system, and $D_{10}$ is the distance to the source in units of 10 kpc.
The resulting inner radius of the accretion disk seems to be considerably close to the compact object in both the analyzed spectra, i.e., on the order of 30$\pm$10 km and 38$^{+14}_{-13}$ km for Spectrum 1 and 2, respectively.\\
These values of the inner radius of the accretion disk are in agreement with those estimated by \cite{zhang_2014}. As suggested by these authors, since the disk is so close to the NS surface it is possible that the electron corona Compton-scatters a fraction of the photons emitted by the innermost part of the accretion disk in addition to the photons emitted by the NS surface. \cite{zhang_2014} tested this hypothesis by adopting a double Comptonization model to fit the persistent spectrum of XB 1916-053. Considering the possibility of this phenomenology, as a further step we reapplied this model to the spectrum already studied by \cite{zhang_2014} (i.e., Spectrum 1), and for the first time to the composite high statistic Spectrum 2.

\begin{figure*}
\centering
\includegraphics[angle=-90, width=8.5cm]{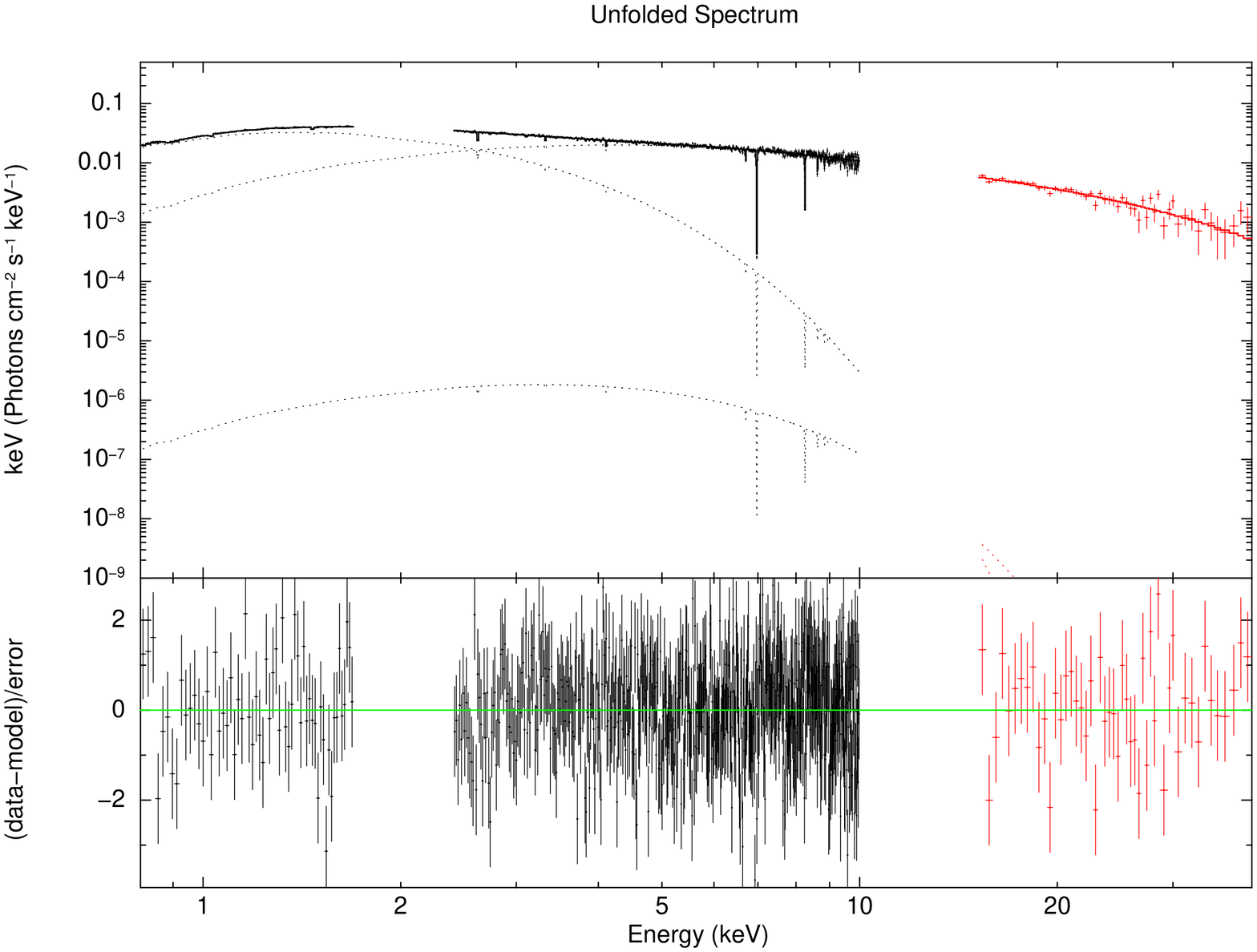}\hspace{0.2truecm}
\includegraphics[angle=-90, width=8.5cm]{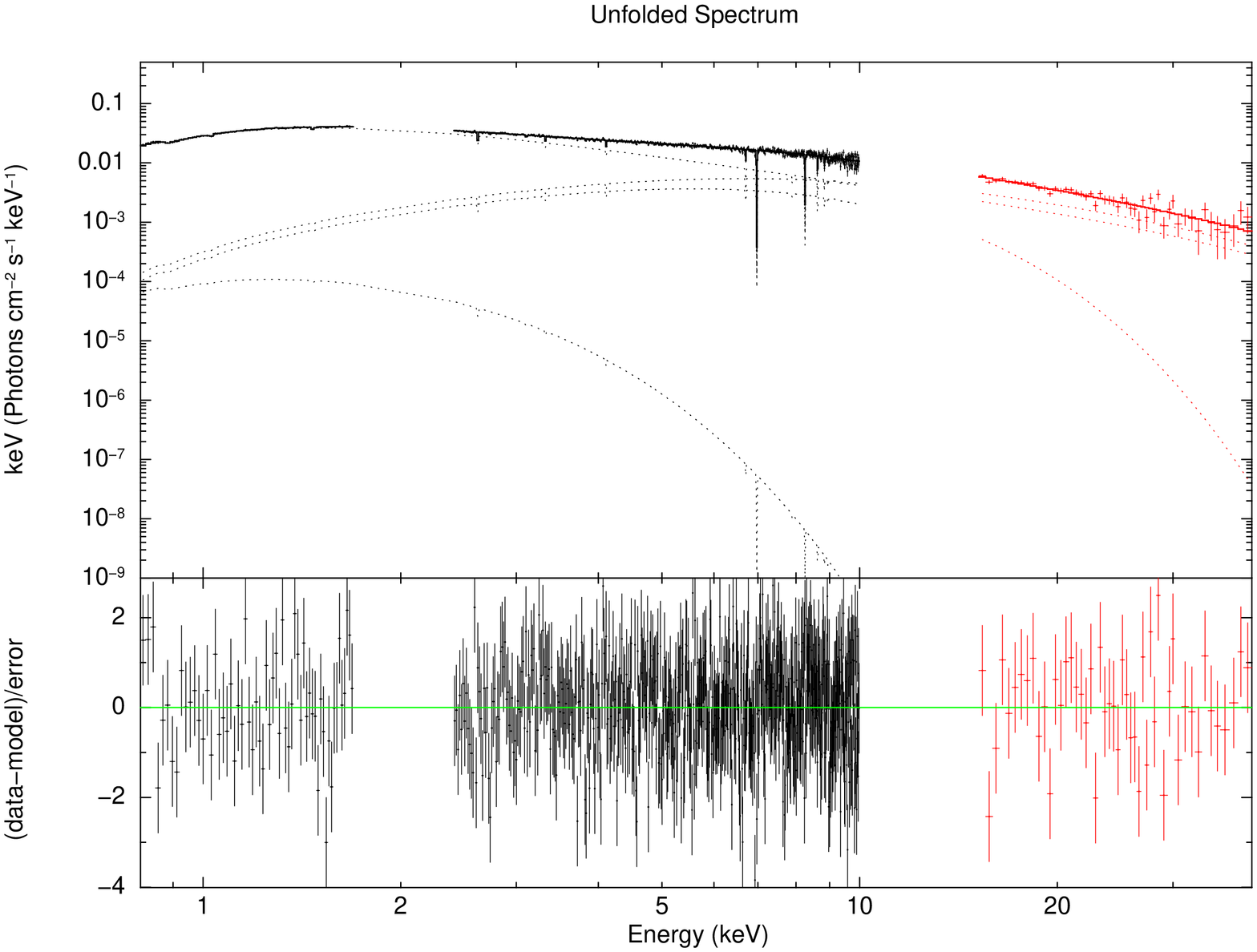}
\caption{Best-fit models performed on data of ObsID 401095010 (Spectrum 1) taking into account the contribution of a local absorber. Left: warm\_1seed model. Right: warm\_2seed model. In each plot the black data represent the XIS03 spectrum, whilst the red data represent the PIN spectrum. The bottom panel of each plot represents the residuals in units of sigma with respect to the adopted model}
\label{fig:spec1_fits}
 \end{figure*}

\begin{figure*}
\centering
\includegraphics[angle=-90, width=8.5cm]{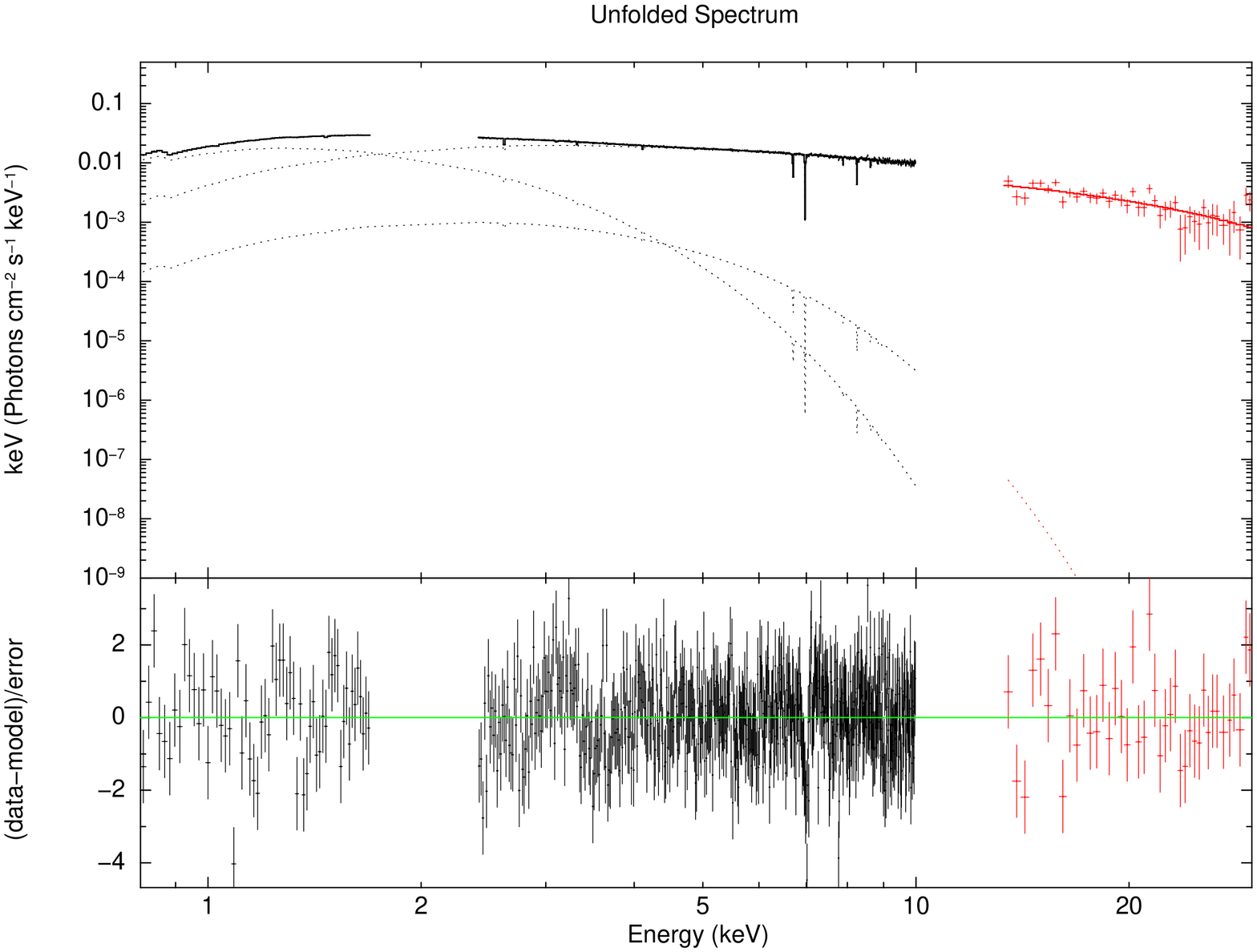}\hspace{0.2truecm}
\includegraphics[angle=-90, width=8.5cm]{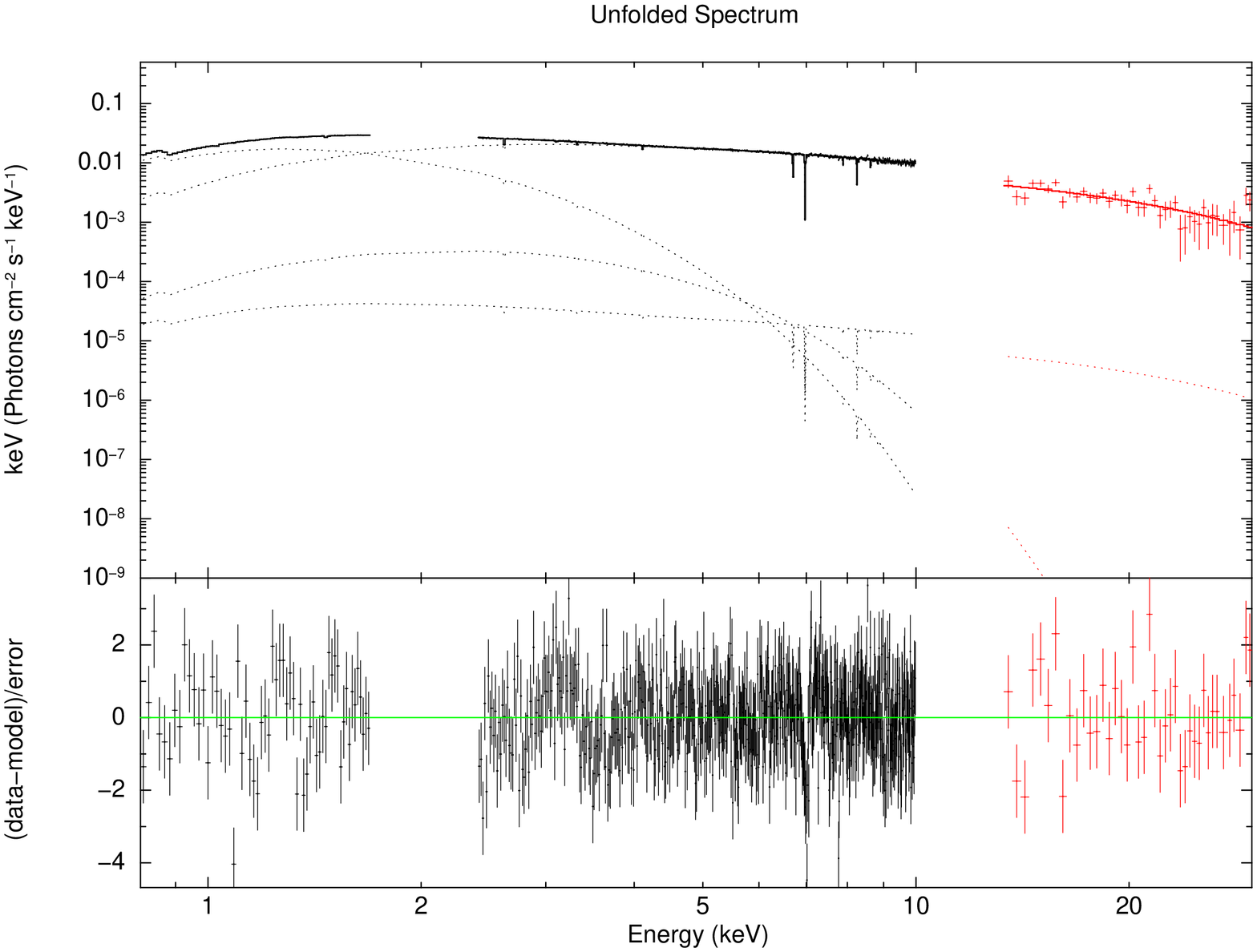}
\caption{Best-fit models performed on data of ObsID 409032010 + 409032020 (Spectrum 2) taking into account the contribution of a local absorber. Left: warm\_1seed model. Right: warm\_2seed model. In each plot the black data represent the combined XIS03 spectra of the two observations, whilst the red data represent the combined PIN spectra of the two observations. The bottom panel of each plot represents the residuals in units of sigma with respect to the adopted model.}
\label{fig:spec2_fits}
 \end{figure*}

The model consists of a further thermal Comptonization component with respect to the case of the 1-seed model.  While the first component (nthComp[inp\_type=0]) scatters photons distributed following a blackbody law, this second component (nthComp[inp\_type=1]) scatters photons that are emitted by the accretion disk and are distributed following a multicolor-disk blackbody law. Assuming that all these photons are scattered by the same electron cloud, we kept tied the values of the photon index $\Gamma$ and electron temperature $kT_e$ for these two Comptonization components. On the other hand, we imposed that the temperature $kT_{bb}$ for (nthComp[inp \_type=0] and (nthComp[inp\_type=1] was the same of the {\tt bbody} and {\tt diskbb} components, respectively.
We also included all the absorption lines we detected using the 1-seed model, leaving the associated energies free to vary and keeping fixed their width to the same value adopted in the first part of the data analysis, that is $\sigma$=20 eV.\\
Fitting this model (hereafter 2-seed model) on Spectrum 1 we obtained a good accordance with the results provided by using the 1-seed model. We extrapolated a total unabsorbed flux of $(5.8\pm0.6)\times 10^{-10}$ erg cm$^{-2}$ s$^{-1}$ in the energy band 0.1-100 keV. The {\tt bbody} component contributes less than 17\% to the total flux, while the {\tt diskbb}, {\tt nthComp[inp\_type=0]} and {\tt  nthComp[inp\_type=1]} components contribute less than 16\%, 52\%, and 59\%, respectively (see the best-fit parameters in Table \ref{tab1}).

We confirm the detection of the K$\alpha$ absorption lines of \ion{Fe}{xxv} and \ion{Fe}{xxvi} with a level of statistical confidence of 4$\sigma$ and 10$\sigma$, respectively.\\
The temperature of the seed photons emitted from the NS surface is of about $kT\sim$ 2 keV, while that related to the accretion disk is of about  $kT_{bb}$=0.6 keV. In addition, the fit returns a photon index for the corona that is equal to $\Gamma$=2.21 and a lower limit on the electron temperature of $kT_e >1$ keV. These parameters are all in agreement with those found using the 1-seed model to fit the spectrum and support the hypothesis that the source could probably be in a soft spectral state, as already suggested by \cite{zhang_2014}.
The fit, however, returned a value of the $\chi^2(d.o.f.)$ of 698(616), which does not represent a statistically significant improvement with respect to the 1-seed model because of the F-test probability that the improvement of the fit occurs by chance on the order of the 5\% .\\ 
We also considered the case in which the corona that envelopes the NS has a different optical depth (and then a different $\Gamma$ index) with respect to the corona that covers the inner regions of the accretion disk. However, we imposed that these coronae have the same electron temperature ($kT_e$). The best-fit parameters are fully in agreement with those obtained with the previous model and we obtained a $\chi^2(d.o.f.)$ of 696.5(615) with an F-test probability of chance improvement of about the 3\%. With a such a poor outcome for the F-test, we did not consider the improvement to the fit statistically significant and we therefore decided not to take a double-$\Gamma$ index corona into account.

Applying the 2-seed model to the higher statistics Spectrum 2 as well, we obtained results that are consistent with those obtained fitting this model onto the Spectrum 1, suggesting that the continuum emission of XB 1916-053 is compatible in the two observations that are separated by a time interval of eight years.\\
We extrapolated a total unabsorbed flux of (5.4$\pm$0.5)$\times 10^{-10}$ erg cm$^{-2}$ s$^{-1}$ in the energy band 0.1-100 keV. For this value of flux the bbody component contributes for less than the 56\%, while the diskbb component contributes about  24\% of the flux and the nthComp[inp\_stype=0] and nthComp[inp\_type=1] components less than 74\% each, respectively.
We also detected the K$\alpha$ absorption lines of \ion{Fe}{xxv} and \ion{Fe}{xxvi} in this case with a level of statistical confidence of 10 and 17$\sigma$, respectively. The K$\beta$ absorption lines of \ion{Fe}{xxv} and \ion{Fe}{xxvi}, on the other hand, were detected with a level of confidence of about 4$\sigma$. \\
Similar to the case in which we fitted the 1-seed model to Spectrum 2, in this case we detected an edge at 0.877 keV that is compatible with a \ion{O}{viii} K absorption edge.
The fit, however, returned a $\chi^2 (d.o.f.)$ of 693.5(599) that actually does not represent a significant improvement of the fit performed on the Spectrum 1 with the same model, since the F-test probability of chance improvement is of about 99\%. Moreover, this fit does not represent an improvement with respect to the 1-seed model applied to the same spectrum, since the F-test probability of chance improvement is about 77\%. \\
As already considered for the Spectrum 1, we tried to take into account different values of the index $\Gamma$ for the two Comptonization components, assuming the same electron temperature. Also in this case, the fit was not sensitive to the change into the spectral model and returned a $\chi^2(d.o.f.)$ of 693.5(598) with an F-test probability that the improvement occurs by chance of 100\% with respect to the previous fit.\\

\begin{table}
\caption{Results of the fit on the dip spectra of Obs 2 and Obs 3, adopting the 1-seed model together with the contribution of two local warm absorbers, respectively. The errors for each spectral parameter are quoted at  90\% of statistical confidence. }
\footnotesize

\begin{tabular}{llcc}
\hline
Model & Component & Obs 2 & Obs 3 \\
{\sc phabs} & nH(10$^{22}$) & 0.479 (frozen) & 0.479 (frozen) \\
{\sc cabs} & nH(10$^{22}$) & $15^{+2}_{-3}$ & $16\pm1$ \\
{\sc cabs} & nH(10$^{22}$) & $67\pm4$ & $91^{+6}_{-4}$  \\
{\sc zxipcf} & Nh(10$^{22}$) & $15^{+2}_{-3}$ & $17\pm1$ \\
 & log$_{xi}$ & $0.7^{+0.5}_{-0.6}$ & $1.14^{+0.14}_{-0.3}$\\
 & CvrFract & $0.80\pm0.03$ & $0.891^{+0.010}_{-0.012}$ \\
 & Redshift & 0 (frozen) & 0 (frozen) \\
{\sc zxipcf} & Nh(10$^{22}$) & $67\pm4$ & $91^{+6}_{-4}$\\
 & log$_{xi}$ & $3.24^{+0.2}_{-0.13}$ & $3.05\pm0.12$ \\
 & CvrFract & $0.64^{+0.12}_{-0.13}$ & $0.57^{+0.09}_{-0.10}$ \\
 & Redshift & 0 (frozen) & 0 (frozen) \\
{\sc bbody} & kT(keV) & 0.75 (frozen) & 0.75 (frozen)\\
 & norm(10$^{-4}$) & 5.0 (frozen) &  5.0 (frozen) \\
{\sc diskbb} & Tin(keV) & 0.51 (frozen) & 0.51 (frozen) \\
 & norm & 78 (frozen) & 78 (frozen) \\
{\sc nthComp} & Gamma & 1.84 (frozen) & 1.84 (frozen) \\
 & kT$_{e}$(keV) & 6 (frozen) & 6 (frozen)\\
 & kT$_{bb}$(keV) & 0.75 (frozen) & 0.75 (frozen) \\
 & inp$_{type}$(0/1) & 0 (frozen) & 0 (frozen) \\
 & Redshift & 0 (frozen) & 0 (frozen)\\
 & norm & 0.01 (frozen) & 0.01 (frozen) \\
\hline
 & $\chi^2/dof$ & 564.8/531 & 554.8/541 \\
\end{tabular}
\label{tab3}
\end{table}

From a comparison between the two adopted models, it is possible to notice a certain indetermination in the evaluation of the goodness of the 2-seed model with respect to the single seed model. The F-test probability that the improvement introduced by the adoption of the 2-seed (with respect to the 1-seed model) occurs by chance is equal to  1\% in the case of the lower statistic Spectrum 1 and to  77\% for Spectrum 2. \\
The presence of several absorption lines due to the presence of different iron ions, however, suggests the possibility that a local partially ionized absorber is located between the observer and the system, as already suggested by \cite{Boirin}. To verify this scenario, we modified the 1-seed and 2-seed models adding the multiplicative spectral component {\tt zxipcf} and getting rid of the Gaussian components used to model the absorption lines and the edge. Hereafter, we call warm\_1seed model the 1-seed model to which we applied the zxipcf component, while the application of the same component to the 2-seed model results in a model that is named warm\_2seed model. \\
The {\tt zxipcf} component uses a grid of XSTAR photoionized absorption models to describe the absorption of the incoming radiation by the plasma, taking into account a micro-turbulent velocity of the plasma of 200 km/s. The model assumes that the absorbing plasma covers only a fraction of the source, whilst the remaining part of the spectrum is seen directly \citep[see ][]{Reeves_2008, Miller_2007}. Moreover, this model  takes into account the column density of absorbing material $N_H$, the ionization state of the plasma $log(\xi),$ and the covering fraction and redshift. The results of the fits reported in Table \ref{tab2}, show similar and compatible values with respect to those obtained with the 1-seed and 2-seed models. \\
The unabsorbed flux extrapolated in the band 0.1-100 keV by fitting Spectrum 1 and Spectrum 2 with the model warm\_1seed is equal to $(6.0\pm 0.6)\times 10^{-10}$ erg cm$^{-2}$ s$^{-1}$ and $(5.1\pm 0.5)\times 10^{-10}$ erg cm$^{-2}$ s$^{-1}$, respectively. We extrapolated the same fluxes adopting the warm\_2seed model to the same observations (see Table \ref{tab2}). \\
Adopting the warm\_1seed model, the contribution of the {\tt blackbody} component is quite small, since this component contributes with less than 7\% and 8\% to the total flux in Spectrum 1 and Spectrum 2, respectively. In addition, the {\tt diskbb} component contributes 33\% and 22\% of the flux, while the {\tt nthComp[inp\_type=0]} contributes 67\% and 77\% for Spectrum 1 and Spectrum 2, respectively.  On the other hand, adopting the warm\_2seed model we notice that the {\tt blackbody} component contributes with less than 17\% and 4\% for Spectrum 1 and Spectrum 2, while the {\tt diskbb} component contributes with less than 10\% for Spectrum 1 and 22\% Spectrum 2. Moreover, the {\tt nthComp[inp\_type=0]} contributes with less than 50\% for Spectrum 1 and 79\% for Spectrum 2, while the {\tt nthComp[inp\_type=1]} component contributes 67\% of the flux for Spectrum 1 and less than18\% for Spectrum 2. \\
The addition of the absorption component due to a warm partially ionized absorber well fits the absorption line shapes and returns values of the $\chi^{2}$(d.o.f.) of 698.3(618) and 707.51(605) for the warm\_1seed model applied to Spectrum 1 and Spectrum 2, respectively, and of 691.3(617) and 707.48(604) for the warm\_2seed model applied to Spectrum 1 and Spectrum 2, respectively. Also in this case, the improvement introduced by the adoption of the warm\_2seed model with respect to the warm\_1seed is not statistically significant, owing to the F-test probability of chance improvement that is equal to 1\% and 87\% for Spectrum 1 and Spectrum 2, respectively; this is exactly what occurred in the case of the 1-seed and 2-seed models. According to the case of the Spectrum 2, this result could actually reflect the poor contribution of the nthComp[inp\_type=1] component in terms of flux (the <18\% of the total flux) with respect to the nthComp[inp\_type=0] component, which provides a higher contribution of flux ( 79\% of the total flux).
In Fig. \ref{fig:spec1_fits} and \ref{fig:spec2_fits} we show the best-fit models warm\_1seed and warm\_2seed applied to Spectrum 1 and Spectrum 2, respectively, with the relative residuals.  \\
Even though it is not possible to break the indetermination of the adopted models,  in the following we discuss the parameters obtained by fitting the warm\_1seed model on Spectrum 2, that is the spectrum with higher statistics, preferring a less parameterized model.

\subsection{Dip spectrum}

While the persistent emission spectrum of XB 1916-053 does not show spectral changes between the Obs 2 and Obs 3, on the other hand, the dip spectrum appears to be inconsistent between these two observations, in particular at the low energies. As a consequence of this, we performed an individual analysis to the dip spectra of these two observations. \\
Since the dip is caused by a gradual covering of the central emitting area of the binary system, resulting in a progressive photoelectric absorption of the incoming flux, we used only the spectra extracted from the XIS data between 0.7 and 10 keV where the dip is more evident.\\
According to the nature of the dipping phenomenon, we modeled the spectra of Obs 2 and Obs 3 using the warm\_1seed model adopted for the continuum emission.\\

The spectral parameters that describe the continuum persistent emission, including the value of the equivalent hydrogen column of the {\tt phabs} component related to the ISM, were kept fixed to those obtained using the warm\_1seed model onto Spectrum 2; an exception to this was the component {\tt zxipcf}. We also took into account that the warm absorber could also be responsible for a continuum optically thin Compton scattering, and for this reason we included a model of optically thin Compton scattering ({\tt cabs} in Xspec), keeping the value of the hydrogen column of this component tied to that of the warm absorber. With this model the complex of features at about 6.7 keV already detected in the persistent spectrum was not fitted; this resulted in a poor value of the $\chi(d.o.f.)$ of 787.44(534) and 1167.9(544) for Obs 2 and Obs 3, respectively. For this reason, we added a further {\tt zxipcf} component, testing the possibility that the local partially ionized absorber showed a nonuniform value of the $log(\xi)$ parameter. Hereafter this model is named 2warm\_1seed model. The best-fit parameters obtained from the fits are reported in Table \ref{tab3}.  \\ 

The spectral parameters of Obs 1 and Obs 2 are quite similar and the components with the larger hydrogen columns $N_H$ are characterized by the higher value of $log(\xi)$ and smaller covering fraction. The most ionized component shows an average value of $log(\xi)$ of about 3.2 in the two observations that well fits the residuals in absorption at about 6.7 keV. We obtain a $\chi^2(d.o.f.)$ of 564.8(531) and 554.8(541), for Obs 2 and Obs 3, respectively. This results in a significant statistical improvement of the fit with respect to the previous fit because the F-test probability that the improvements occur by chance is about $5\times10^{-38}$ and $5\times10^{-87}$, respectively.

\begin{figure}
\centering
\includegraphics[angle=-90, width=8.5cm]{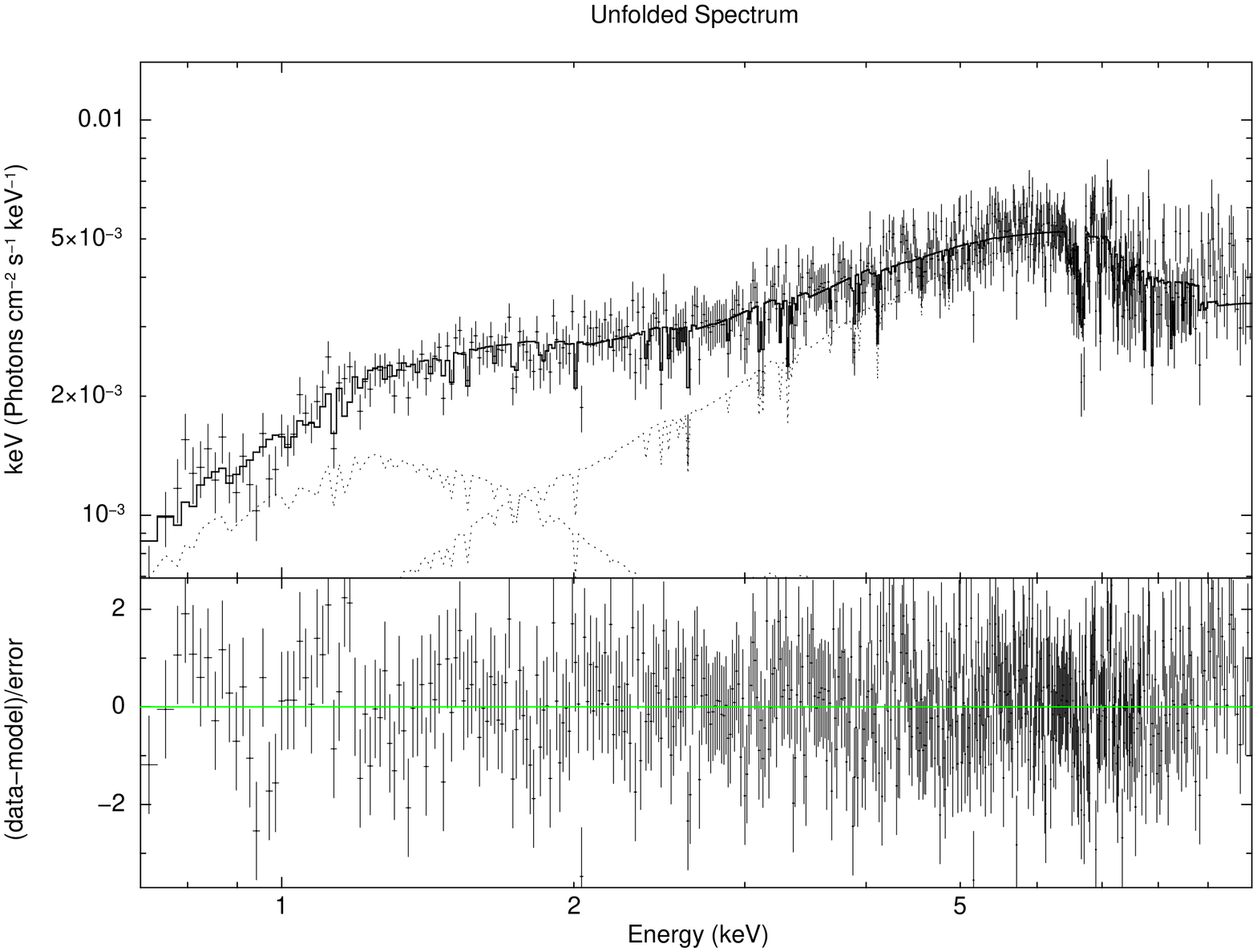}\hspace{0.2truecm}
\includegraphics[angle=-90, width=8.5cm]{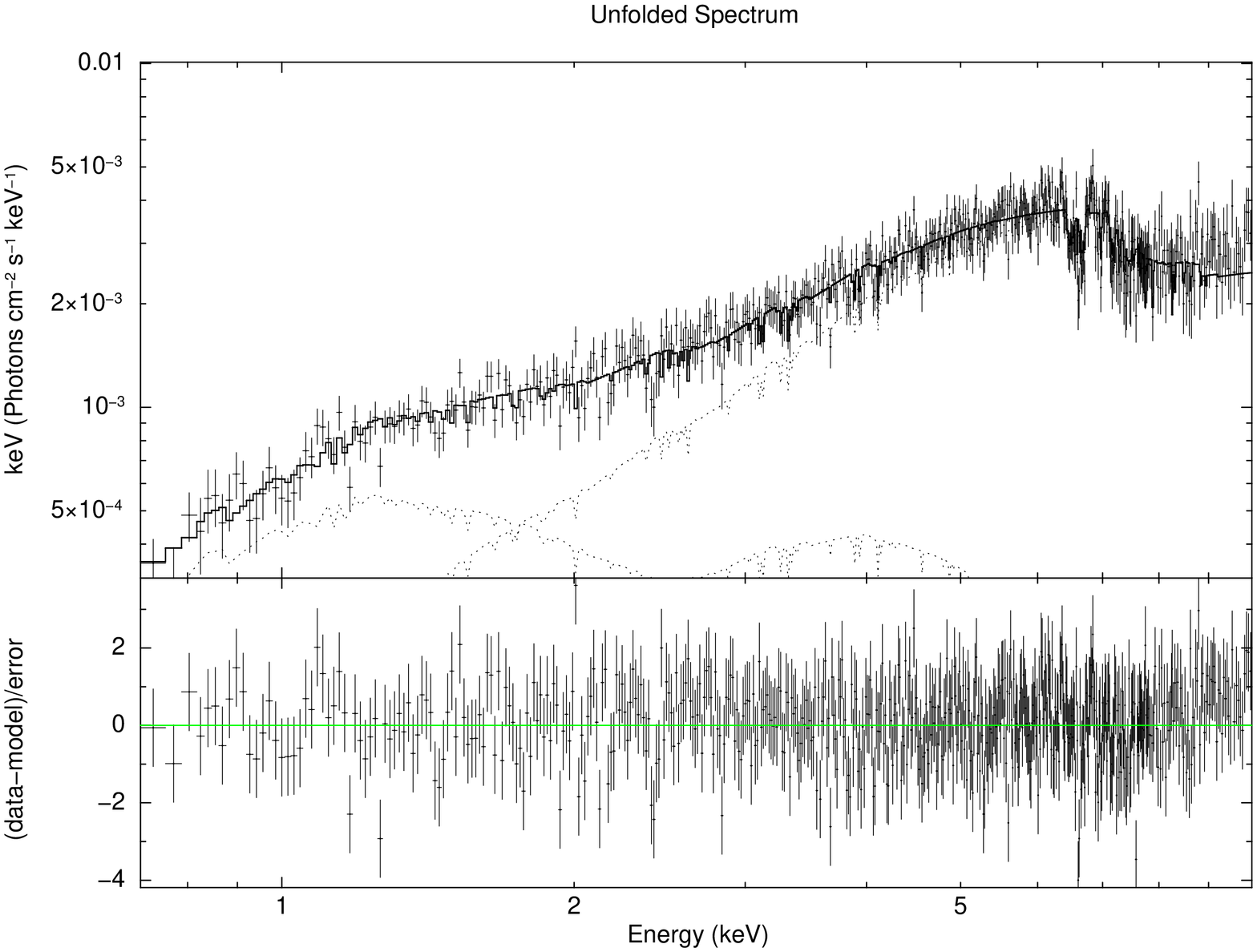}
\caption{Best-fit model for the dip spectra obtained from the Suzaku/XIS03 data of ObsID 409032010 (Obs2, upper plot) and ObsID 409032010 (Obs3, bottom plot). The residuals with respect to the adopted model are shown in units of sigma in the bottom panel of each plot.}
\label{fig:spec_dip}
 \end{figure}

\section{Discussion}
\label{sec:discussion}

\subsection{Persistent spectrum}

We analyzed and modeled the continuum and dip spectra of the LMXB source XB 1916-053, using high statistics observations collected by the Suzaku space mission at different times. \\
Fitting the spectra of the persistent emission with the 1-seed model, we observed absorption features that suggested the presence of a local partially ionized absorber. For this reason, we adopted the warm\_1seed model that includes the contribution of the warm partially ionized local absorber {\tt zxipcf}.
Fitting the warm\_1seed model to the Spectrum 1, we obtained spectral parameters that are in good agreement with the results of \cite{zhang_2014}. \\
The fit returns a value of the photon index of $\Gamma$=2.1$^{+0.3}_{-0.2}$ and a temperature of the Comptonizing cloud of $kT_e =7\pm1$ keV. 
Moreover, the normalization parameter of the {\tt diskbb} component allowed us to estimate the apparent inner radius of the accretion disk as $R_{in}=32\pm10$ km, conveniently corrected for the square of the color factor of \cite{Shimura_95}. This estimation assumed an inclination angle of $i=70 \pm 10 \deg$ \citep[][]{Van_Parad_1988, Frank_1987} and a distance of $D=9.3$ kpc as evaluated by \cite{Yoshida_PhD_93} with an associated error of 20\% \citep[][]{zhang_2014}.
Such value of the inner radius is in agreement with other radius measurements performed in other atoll sources \citep[e.g.,][]{disalvo_15} and suggests that the accretion disk extends close to the NS surface, as observed by \cite{zhang_2014} who analyzed the same observation. 
The application of the same model to Spectrum 2 returns similar results. The values of the photon index and temperature of the electron cloud ($\Gamma = 1.84^{+0.04}_{-0.05}$ and $KT_e=6\pm1$ keV, respectively) are compatible with those obtained from the previous fit, suggesting that no evident changes in the corona have been observed in the two observations that are separated by a time span of eight years. 
In this case we infer an inner radius for the accretion disk of $41^{+15}_{-13}$ km that is in agreement with the previous estimation. \\ 
We extracted the unabsorbed bolometric flux between 0.1 and 100 keV for each spectrum modeled with the warm\_1seed model, obtaining $\Phi=(5.8\pm 0.6)\times 10^{-10}$ erg cm$^{-2}$ s$^{-1}$ and $\Phi=(5.4\pm 0.5)\times 10^{-10}$ erg cm$^{-2}$ s$^{-1}$ for Spectrum 1 and Spectrum 2, respectively. Assuming a distance to the source of $d=9.3 \pm 1.4$ kpc \citep{Yoshida_PhD_93}, we obtained an unabsorbed bolometric luminosity in the 0.1--100 keV energy band of $L=(6\pm2) \times 10^{36}$ erg s$^{-1}$ and $L=(5\pm2) \times 10^{36}$ erg s$^{-1}$ for Spectrum 1 and Spectrum 2, respectively. This suggests that XB 1916-053 is a persistent source that does not show important changes of flux as also shown by the \textit{All-Sky Monitor (ASM)} on board the \textit{Rossi X-ray Timing Explorer (RXTE)} \citep{Homer_01}.\\

We also evaluated the optical depth of the Comptonizing electron cloud for Spectrum 1 and Spectrum 2, using the relation of \cite{Zdziarski_1996}

\begin{equation}
\label{eq:Gamma_tau}
\Gamma=\left[ \dfrac{9}{4}+\dfrac{1}{\tau \left(1+\dfrac{\tau}{3}\right) \left(\dfrac{kT_e}{m_e c^2}\right)}\right]^{1/2} - \dfrac{1}{2},
\end{equation}
according to the spectral parameters obtained by fitting the warm\_1seed model onto the data.\\
We obtained an optical depth of $\tau =6^{+1}_{-2}$ for Spectrum 1 and $\tau = 8\pm1$ for Spectrum 2, respectively, indicating that the corona is optically thick. \\
An estimation of the radius of the region from which the seed photons are emitted, can be obtained using the relation of \cite{Zand_99}, assuming a spherical geometry of the corona,

\begin{equation}
\label{eq:R_seed_sph}
R_0=3\times 10^4 \; d \left( \dfrac{f_{bol}}{1+y} \right)^{1/2} \left(kT_{bb}\right)^{-2},
\end{equation}
where $d$ is the distance to the source in kpc, $f_{bol}$ is the unabsorbed bolometric flux extrapolated from the Comptonization component in erg cm$^{-2}$ s$^{-1}$, $kT_{bb}$ is the temperature of the seed photons in keV, and $y=4kT_e max[\tau , \tau ^2]/(m_e c^2)$ is the Compton parameter, in which $kT_e$ is the electron temperature in keV.\\
On the basis of the results we obtained fitting the warm\_1seed model onto the data, we calculated a Comptonization parameter $y=0.6^{+0.2}_{-0.4}$ and $y=3\pm1$ for Spectrum 1 and Spectrum 2 and a radius for the seed photons of $R_0=10^{+2}_{-3}$ km and $R_0= 14^{+6}_{-3}$ km, respectively, already corrected for the color factor $\psi = 1.7$ of \cite{Shimura_95}. This value agrees with the typical radii of NSs and suggests that the emission from the NS surface actually contributes to provide photons that are Compton-scattered by the corona. This result is also consistent with the quite low value of normalization found for the {\tt bbody} component. The direct emission of the NS surface, indeed, could be totally intercepted by the optically thick corona, then contributing as source of seed photons.\\
The obtained optical depth and the values found for the temperature of the electron cloud seem to suggest that in the two spectra the physical conditions of the corona did not change and that the accretion disk maintains its inner radius close to the NS surface, according to what is usually observed for systems in a soft spectral state. \\

In their work, \cite{zhang_2014} suggested that the 2-seed model better describes the Comptonization process that contributes to the observed spectrum of the source. Furthermore, in \cite{Zhang_16} the persistent spectrum of the dipping and eclipsing LMXB source EXO 0748-676 is modeled with a 2-seed model on the basis of the fact that the authors obtained a radius of the seed photons that is larger than the radius of the NS, and then a contribution of photons is required for the Comptonization. This is not the case for our data, for which the adoption of the warm\_1seed model implies that the photons that are Compton-scattered in the corona are emitted in a region with a radius that is compatible with the radius of the NS. However, to test if the contribution of the photons emitted by the disk is actually negligible, in Section \ref{sec:analysis} we decided to model our spectra with the warm\_2seed model as well.  \\
The results obtained with this model, reinforced by the F-test, suggested that a double Comptonization process is not needed on the basis of the current statistics of the data, while the fit seems to suggest an inner radius of the accretion disk that extends relatively close to the NS surface. \\

Adopting the warm\_2seed model and using Eq. \ref{eq:Gamma_tau}, we obtained an optical depth of $\tau=3^{+8}_{-2}$ and $\tau=8^{+5}_{-3}$ for Spectrum 1 and Spectrum 2, respectively. Using these optical depths and the electron temperature $kT_e$ obtained from the fit (see Table \ref{tab2}) we found a Comptonization parameter $y=1^{+5}_{-1}$ and $y=3^{+4}_{-2}$, for Spectrum 1 and Spectrum 2, respectively.\\

In this case, using Eq. \ref{eq:R_seed_sph} we inferred that the radius of the region from which the seed photons are emitted is $R_0$ <3 km for Spectrum 1, while it is equal to $R_0 = 5^{+6}_{-2}$ km for Spectrum 2. The radius obtained for Spectrum 1 is fairly small, if compared to the typical value of the NS radius (i.e., $\sim$ 10 km). This discrepancy could be caused by a low statistics of the Spectrum 1 or could alternatively suggest that only a fraction of the NS surface is responsible for the emission of seed photons that are Compton scattered. On the other hand, the higher statistics of Spectrum 2 allowed us to better constrain the value of $R_{0}$ that is in agreement, within the error, with the typical value of NSs radii. \\
Taking into account all these results, we are not able to break the degeneracy of the adopted models, even through the analysis of a spectrum with higher statistics. For this reason we take into account the results of the warm\_1seed model, which not only returns a better value of the $\chi^2$ but is also less parameterized.\\

The presence of a local partially ionized absorbing material was justified by the observation of several absorption features in the spectra. Fitting Spectrum 1 and Spectrum 2 with the 1-seed model we observed the absorption K$\alpha$ lines of \ion{Fe}{xxv} and \ion{Fe}{xxvi}. These lines are quite strong and were detected with a level of statistical confidence of 4$\sigma$ and 11$\sigma$ for the \ion{Fe}{xxv} and \ion{Fe}{xxvi} K$\alpha$ lines, respectively, in the case of Spectrum 1, and with a level of confidence of 9$\sigma$ and 17$\sigma$ for Spectrum 2. 
\cite{Iaria_06} reported that these features originate at the outer rim of the accretion disk, where the plasma is colder but also partially ionized. The strength of these lines, and in particular the difference in the line depth, can be explained as a difference of the number of the \ion{Fe}{xxv} ions with respect to \ion{Fe}{xxvi}, for the K$\alpha$ transitions. To explain the larger depth of the \ion{Fe}{xxvi} absorption line, we should expect a prevalence of these ions with respect to the \ion{Fe}{xxv} ions.
To estimate the difference of population for the two ions, we used the relation provided by \cite{Spitzer_1978}, i.e.,

\begin{equation}
\label{eq:population}
\dfrac{W_{\lambda}}{\lambda}=\dfrac{\pi e^2}{m_e c^2}N_j \lambda f_{ij},
\end{equation}
where $N_j$ is the column density for the relevant species, $f_{ij}$ is the oscillator strength, $W_{\lambda}$ is the equivalent width of the line, and $\lambda$ is the wavelength expressed in centimeters. \\

The values of the oscillator strength are tabulated in \cite{Verako95} and are equal to $f_{ij}= 0.798$ and $f_{ij}= 0.277$ for the \ion{Fe}{xxv} and \ion{Fe}{xxvi} K$\alpha$ transitions, respectively. 
From the 1-seed model applied to Spectrum 2 we find an equivalent width of $W_{\lambda}= (5\pm 2)$ eV and $W_{\lambda} =(17\pm2)$ eV for the \ion{Fe}{xxv} and \ion{Fe}{xxvi} K$\alpha$ absorption lines, respectively. The errors associated with the equivalent widths are reported with a level of statistical confidence of 1$\sigma$. Then, using Eq. \ref{eq:population} we obtain $N_{Fe25,k\alpha}= (0.6\pm0.2) \times 10^{17}$ cm$^{-2}$ and $N_{Fe26,k\alpha}= (5.6\pm 0.7) \times 10^{17}$ cm$^{-2}$, whose ratio suggest a predominance of \ion{Fe}{xxvi} ions with respect to the \ion{Fe}{xxv} by a factor of about 9, and this probably explains the difference in the depth of the two lines.
These strong absorption lines were already noticed by \cite{zhang_2014}, \cite{Boirin_xmm}, and \cite{Iaria_06}, where a full diagnostics of the absorption lines was performed using Chandra data.\\
The Spectrum 2, however, has a higher statistics with respect to the Spectrum 1 and in addition to the K$\alpha$ absorption lines it also shows residuals that we found to be consistent with the K$\beta$ absorption lines of \ion{Fe}{xxv} and \ion{Fe}{xxvi}, respectively. \\
Because of the moderate energy resolution of Suzaku, these two absorption lines have been modeled keeping their widths fixed to the same value of the K$\alpha$ lines, i.e., 20 eV. The inclusion of these two lines in the 2-seed model actually improves the fit with respect to the case in which only the K$\alpha$ lines were included because the probability for such  improvement to occur by chance is $2\times 10^{-3}$.\\

The presence of the \ion{Fe}{xxv} K$\beta$ line has been suggested by \cite{Boirin_xmm} with XMM-Newton data, and is confirmed in this work taking advantage of the statistics of Spectrum 2. We detected this line with a level of confidence of 3$\sigma$ and about 4$\sigma$ using the 1-seed and 2-seed models, respectively. Moreover, we also detected a line consistent with a \ion{Fe}{xxvi} K$\beta$ absorption line with a level of confidence of about 4$\sigma$, independently from the adopted model. 
Actually, this line is located in the energy range in which \cite{Boirin_xmm} observed some weak residuals that assumed to be caused by the presence of \ion{Ni}{XXVII}. We introduce the further possibility that this feature is due to the presence of \ion{Fe}{xxvi} ions making K$\beta$ transitions. \\
Taking into account Eq. \ref{eq:population}, we can estimate the number population of the two species of ions. From the 1-seed model fitted to Spectrum 2 we obtain an equivalent width of $W_{\lambda}=(5\pm 3)$ eV for both the \ion{Fe}{xxv} and \ion{Fe}{xxvi} K$\beta$ transitions, respectively, using the equivalent oscillator strengths $f_{ij}=0.156$ and $f_{ij}=0.079$, respectively, tabulated in \cite{Verako95}. We obtain $N_{Fe25,k\beta}= (3\pm 2) \times 10^{17}$ cm$^{-2}$ and $N_{Fe26,k\beta}= (6\pm 3) \times 10^{17}$ cm$^{-2}$, with a ratio that suggests a predominance by a factor of about 2 of \ion{Fe}{xxvi} ions making k$\beta$ transitions with respect to \ion{Fe}{xxv} ions making the same transition.\\
In addition to these discrete features, we also detected an absorption edge in Spectrum 2 at 0.87 keV. This edge is compatible with a \ion{O}{viii} K absorption edge that has already been observed in several LMXB systems \citep[see, e.g.,][]{Iaria_1702_2016, Cottam_2001}. This feature is necessary to fit Spectrum 2 with all the adopted models. In particular, taking into account that the local warm absorbed in the system has a considerably high ionization state (as suggested by the $log(\xi)$ value of about 5), it is probable that the \ion{O}{viii} ions that are responsible for this feature are generated in a relatively colder region.

\subsection{Dip spectrum}

The dip spectrum, on the other hand, was studied considering the XIS data of the ObsID 409032010 (Obs 2) and 409032020 (Obs 3). The resulting spectra have a net exposure of about 26 ks and 46 ks for Obs 2 and Obs 3, respectively.\\

The results of the fits performed using the 2warm\_1seed model show that the introduction of a further warm absorber into the warm\_1seed model is statistically needed by the available data, which for this reason seem to suggest that the absorber is not uniformly ionized. 
Moreover, even though the estimated covering fractions of the absorbers are similar passing from Obs 2 to Obs 3, on the other hand the absorber provided with the higher value of $log(\xi)$ increases its equivalent hydrogen column density.  

The local absorbing materials have an equivalent hydrogen column density of $N_H=(15^{+2}_{-3})\times 10^{22}$ cm$^{-2}$ and $N_H=(67\pm4)\times 10^{22}$ cm$^{-2}$ in Obs 2, or alternatively of $N_H=(17\pm1)\times 10^{22}$ cm$^{-2}$ and $N_H=(91^{+6}_{-4})\times 10^{22}$ cm$^{-2}$ in Obs 3. The fraction of the persistent spectrum that is covered by this material is on average equal to the 85\% and 60\% for the less and highly ionized absorbers, respectively, in both the observations, while the ionization parameters $log(\xi)$ are on average equal to 0.87 and 3.14 in each observation. These results suggest that the bulge probably does not significantly change its morphology even though the ionization state and density of the matter could show variations in timescales on the order of few orbital periods.\\

A rough estimation of the distance of the warm absorber could be evaluated by means of the ionization parameter $\xi$, which determines the ionization state of the material of the warm absorber. According to \cite{Tarter_69}, this parameter is defined as

\begin{equation}
\label{eq:xi}
\xi = \dfrac{L}{n_{H}r^2},
\end{equation}
where L is the unabsorbed luminosity of the illuminating source in erg s$^{-1}$, $n_{H}$ is the total density of hydrogen in cm$^{-3}$, and $r$ is the distance of the absorber from the center of the system in centimeters. 
Assuming that the radial size of the absorber $\Delta R$ is much lower that the typical value of the outer radius of the accretion disk, that is $r_{disk}=4\times 10^{10}$ cm as reported in \cite{Iaria_06}, we can infer upper limits on the distances of the local absorbers with respect to the position of the NS. We estimate that the less ionized absorber of Obs 2 is located at $r\leq 5\times 10^{11}$cm, while the more ionized at $r\leq 1\times 10^{10}$cm.  In Obs 3 the less ionized absorber is located at $r\leq 3\times 10^{11}$cm, while the more ionized is located at $r\leq 1\times 10^{10}$cm. From these results we can notice that the upper limits on the distances inferred for the more ionized absorber in both observations are identical and actually explain the presence of the absorption features also observed into the persistent spectra, which usually form in the inner parts of the accretion disks. At the same time, these results rescale for the first time the distance of the absorber that \cite{Iaria_06} arbitrarily assumed to be located at the outer rim of the accretion disk. \\ 
At the same time, we are not able to constrain the distance of the weakly ionized absorbers of both observations. For this reason, according to their weak ionization parameters returned by the fits, we assumed we could locate these absorbers at the disk outer rim, trying to estimate their density $n_H$ and thickness $\Delta R$. With this assumption, using the bolometric luminosity of the source in the range 0.1-100 keV (i.e., L=$(5\pm2)\times10^{36}$ erg/s) and the values of $\xi$ we estimate that the total density of hydrogen in the weakly ionized absorber of Obs 2 is of about $n_H\sim7\times10^{14}$ cm$^{-3}$, which implies a thickness of the absorber of $\Delta R \sim 2\times10^{8}$ cm (i.e., about the 0.5\% of the accretion disk radius). On the other hand, for the weakly ionized absorber of Obs 3 we estimate a hydrogen density of $n_H\sim2\times10^{14}$ cm$^{-3}$, which implies a thickness of $\Delta R \sim 8\times10^{8}$ cm (i.e., about the 2\% of the disk radius).

\section{Conclusions}
\label{sec:conclusion}

In this work we report the results of a spectral analysis performed on both the persistent and the dip spectra of the LMXB system XB 1916-053 using Suzaku data. Even though we expect the photons emitted by the accretion disk to be partly scattered by the electron corona, the considerable statistics offered by the most recent observations is not sufficient to discriminate if the Comptonization of these photons is relevant with respect to that of the only photons emitted by the NS surface. We therefore preferred to adopt the single Comptonization description as the best-fit model because in that approach there is less parameterization of the data; however we took the contribution of the local partially ionized absorber into account.

The source has been found in a quite soft spectral state independent of the model adopted to fit the continuum spectrum. The achievement of a greater statistics, on the other hand, allowed us to detect various discrete absorption features, and in particular the K$\beta$ absorption lines of \ion{Fe}{xxv} and \ion{Fe}{xxvi}, in addition to the K$\alpha$ transition lines of the same ions, already reported in literature. The equivalent widths of these absorption lines seem to be consistent with a higher number of \ion{Fe}{xxvi} ions with respect to the \ion{Fe}{xxv}. In addition, we detected a \ion{O}{viii} K absorption edge, which suggests the presence of this ion at large radial distances from the NS (possibly in the outer parts of the accretion disk). \\
The analysis of the dip spectrum highlighted how the dip itself could be explained as a gradual covering of the persistent spectrum by a cold and partially nonuniformly ionized bulk of matter;  according to our results, this dip can be located at a distance from the NS that is lower than $ 1\times 10^{10}$ cm, that is at about  25\% of the disk outer rim according to the existing  estimations in literature. The weakly ionized part of the absorber, on the other hand, is characterized by a total density of hydrogen that slightly variates between $7\times10^{14}$ cm$^{-3}$ and $2\times10^{14}$ cm$^{-3}$ in the two observations and by a thickness of the absorber that variates between 8000 km and 4000 km. In order to better constrain the persistent absorption features and their relative widths, we encourage high exposure observations to be performed by space missions, such as Athena, using high energy resolution.

\section*{Acknowledgements}

This research has made use of data and/or software provided by the High Energy Astrophysics Science Archive Research Center (HEASARC), which is a service of the Astrophysics Science Division at NASA/GSFC and the High Energy Astrophysics Division of the Smithsonian Astrophysical Observatory.\\
This research has made use of the VizieR catalogue access tool, CDS, Strasbourg, France.\\
We acknowledge financial contribution from the agreement ASI-INAF I/037/12/0. We also acknowledge financial contribution from the agreement ASI-INAF n.2017-14-H.0.
We acknowledge support from the HERMES Project, financed by the Italian Space Agency (ASI) Agreement n. 2016/13 U.O.\\
Part of this work has been funded using resources from the research grant “iPeska” (P.I. Andrea Possenti) funded under the INAF national call Prin-SKA/CTA approved with the Presidential Decree 70/2016. S.M.M. thanks the research project “Stelle di neutroni come laboratorio di Fisica della Materia Ultra- densa: uno studio multifrequenza” financed by Regione Autonoma della Sardegna (scientific project manager prof. Luciano Burderi) in which part of this work was developed. \\
The authors also acknowledge the HXD team for the useful help during the data analysis step.


\bibliographystyle{aa}                                          
\bibliography{XB1916}

\end{document}